\documentclass[fleqn,usenatbib,useAMS]{mnras}

\usepackage[dvipsnames]{xcolor}
\usepackage{tikz,hyperref}

\definecolor{lime}{HTML}{A6CE39}
\DeclareRobustCommand{\orcidicon}{
	\begin{tikzpicture}
	\draw[lime, fill=lime] (0,0) 
	circle [radius=0.13] 
	node[white] {{\fontfamily{qag}\selectfont \tiny ID}};
	\draw[white, fill=white] (-0.0625,0.095) 
	circle [radius=0.007];
	\end{tikzpicture}
	\hspace{-2mm}
}

\foreach \x in {A, ..., Z}{\expandafter\xdef\csname orcid\x\endcsname{\noexpand\href{https://orcid.org/\csname orcidauthor\x\endcsname}
			{\noexpand\orcidicon}}
}

\usepackage{newtxtext,newtxmath}
\usepackage{booktabs}
\usepackage[T1]{fontenc}
\usepackage{ae,aecompl}
\usepackage{graphicx}	
\usepackage{amsmath}	
\usepackage{booktabs}
\usepackage{times}
\input{epsf}

\title[R Aqr during periastron passage]{The impact of periastron passage on the X-ray and optical properties of the Symbiotic System R Aquarii}

\author[D.~A.~Vasquez-Torres et al.]{D.~A.~Vasquez-Torres\thanks{E-mail:\,d.vasquez@irya.unam.mx}$^{1}\orcidA$, J.~A.~Toal\'{a}\thanks{Visiting astronomer at the IAA-CSIC as part of the Centro de Excelencia Severo Ochoa Visiting-Incoming programme.}$^{1}\orcidB$, A.~Sacchi$^{2}\orcidC$, M.~A.~Guerrero$^{3}\orcidF$, E.~Tejeda$^{4}\orcidG$, M.~Karovska$^{2}\orcidD$ 
\newauthor and R.~Montez Jr.$^{2}\orcidE$
\\
$^{1}$Instituto de Radioastronom\'{i}a y Astrof\'{i}sica, Universidad Nacional Aut\'{o}noma de M\'{e}xico, 58090 Morelia, Michoac\'{a}n, Mexico\\
$^{2}$Center for Astrophysics | Harvard \& Smithsonian, 60 Garden Street, Cambridge, MA 02138, USA\\
$^{3}$Instituto de Astrof\'{i}sica de Andaluc\'{i}a, IAA-CSIC, Glorieta de la Astronom\'{i}a S/N, Granada 18008, Spain\\
$^{4}$CONAHCYT - Instituto de F\'{i}sica y Matem\'{a}ticas, Universidad Michoacana de San Nicol\'{a}s de Hidalgo, Ciudad Universitaria, 58040 Morelia, Mich., Mexico
}

\date{\today}

\pubyear{2024}

\begin{document}
\label{firstpage}
\pagerange{\pageref{firstpage}--\pageref{lastpage}}
\maketitle

\begin{abstract}
\noindent Multi-epoch {\it Chandra} and {\it XMM-Newton} observations of the symbiotic system R Aquarii (R Aqr) spanning 22 yr are analysed by means of a reflection model produced by an accretion disc. This methodology helps dissecting the contribution from different components in the X-ray spectra of R Aqr: the soft emission from the jet and extended emission, the heavily-extinguished plasma component of the boundary layer and the reflection contribution, which naturally includes the 6.4 keV Fe fluorescent line. The evolution with time of the different components is studied for epochs between 2000 Sep and 2022 Dec, and it is found that the fluxes of the boundary layer and that of the reflecting component increase as the stellar components in R\,Aqr approach periastron passage, a similar behaviour is exhibited by the shocked plasma produced by the precessing jet. Using publicly available optical and UV data we are able to study the evolution of the mass-accretion rate $\dot{M}_\mathrm{acc}$ and the wind accretion efficiency $\eta$ during periastron. These exhibit a small degree of variability with median values of $\dot{M}_\mathrm{acc}$=7.3$\times10^{-10}$~M$_\odot$~yr$^{-1}$ and $\eta$=7$\times10^{-3}$. We compare our estimations with predictions from a modified Bondi-Hoyle-Lyttleton accretion scenario.
\end{abstract}

\begin{keywords}
(stars:) binaries: symbiotic  --- accretion, accretion discs --- X-rays: stars --- X-rays: binaries --- ISM: individual objects: R\,Aquarii ---
ISM: jets and outflows
\end{keywords}

\section{INTRODUCTION}
\label{sec:intro}

R Aquarii (hereinafter R Aqr) is one of the closest symbiotic systems, with an estimated distance $d$ ranging from about $\sim$180 to $\sim$260 pc \citep[][]{Liimets2028, Alcolea2023}, with the farthest value of 387$^{+50}_{-110}$ pc estimated from {\it Gaia} parallax measurements \citep[][]{BailerJones2021}. Its proximity allows astronomers to characterise the symbiotic system and its associated nebula with extreme detail. For decades, optical studies have shown the intricate morphology of the nebula around R Aqr \citep[see][and references therein]{Liimets2028} consisting of (at least) three main structures: an outer hourglass structure encompassing an inner bipolar structure and a spiral-like filament entwined around the latter. It has been argued recently by \citet{santamaria2024} that the action of the precessing jet at the core of R Aqr is responsible for the current morphology of the nebula associated with this symbiotic system.

The most recent orbit calculations of this symbiotic system predict an orbital period of 42.4$\pm$0.2 yr \citep[see, e.g.,][]{Alcolea2023}, an eccentricity of 0.45$\pm$0.01, an inclination of the orbital plane with respect to the line of sight of 110.7$\pm$1.0$^\circ$, and a measured semi-major axis of about 47--57 mas \citep{Alcolea2023,Bujarrabal2018}. 
The periastron is estimated to have occurred during 2019.9$\pm$0.1. R Aqr consists of a white dwarf (WD) with a mass of $M_\mathrm{WD}=0.7 \pm 0.2$ M$_\odot$ and a M-type cool component with a mass of 1.0$\pm$0.2 M$_\odot$. The cool component is a long-period Mira-type variable with a pulsation period of $P$=388.1$\pm$0.1 d \citep[][]{Gromadzki2009}. These parameters translate into a radius for the M-type star of $\sim$250~R$_\odot$ using the period--mass relationship \citep[see][]{Wood1993}.

Several authors have estimated the mass accretion rate ($\dot{M}_\mathrm{acc}$) of the WD component using a variety of methodologies and assumptions, with values in the  $\sim$10$^{-9}$--10$^{-8}$ M$_\odot$~yr$^{-1}$ range \citep[See e.g.][]{Burgarella1992, Henney1992, Ragland2008, Melnikov2018, Sacchi2024}. \citet{Burgarella1992} argued against the presence of an accretion disc around the WD component in R Aqr and the lack of flickering in the optical spectra of this symbiotic system seems to be in line with that result \citep{Snaid2018}. However, based on near-IR observations, \citet{Hinkle2022} suggested that an accretion disc is present and has a size of $\gtrsim$5~AU, with estimations of 4~AU during periastron passage given the elliptical orbit.

\begin{table*}
\begin{center}
\caption{Details of the X-ray observations of R Aqr analysed in this paper.}
\footnotesize
\setlength{\tabcolsep}{\tabcolsep}  
\begin{tabular}{lcccccc}
\hline
Instrument  & Obs.~ID. & Observation date  & Epoch  & Exp. Time  & Useful time &  Net count rate \\
           &         &     (UTC)            & (yr)  & (ks)     & (ks)        &  (cnt~ks$^{-1}$)   \\
\hline
{\it Chandra} ACIS-S & 651    & 2000-09-10T04:58:36 & 2000.69 & 22.72 & 22.72 & 5.8$\pm$0.5 \\
{\it Chandra} ACIS-S & 4546   & 2003-12-31T15:04:37 & 2003.99 & 36.52 & 36.13 & 6.4$\pm$0.4 \\
{\it Chandra} ACIS-S & 5438   & 2005-10-09T19:08:50 & 2005.78 & 66.76 & 66.76 & 8.9$\pm$0.4 \\
{\it Chandra} ACIS-S & 19015  & 2017-10-13T13:48:02 & 2017.79 & 75.66 & 75.46 & 34.5$\pm$0.7 \\
{\it Chandra} ACIS-S & 20809  & 2017-10-11T08:22:21 & 2017.78 & 49.37 & 48.19 & 29.9$\pm$0.8\\
{\it Chandra} ACIS-S & 23108  & 2020-01-11T23:53:56 & 2020.03 & 47.48 & 46.63 & 42.3$\pm$1.0\\
{\it Chandra} ACIS-S & 23325  & 2021-04-29T21:38:04 & 2021.33 & 31.91 & 30.94 & 32.8$\pm$1.0\\
{\it Chandra} ACIS-S & 24341  & 2021-04-30T18:00:36 & 2021.34 & 35.60 & 35.21 & 26.6$\pm$0.8\\
{\it Chandra} ACIS-S & 27322  & 2022-10-03T18:33:27 & 2022.75 & 18.83 & 18.73 & 10.9$\pm$0.8\\
{\it Chandra} ACIS-S & 27333  & 2022-09-26T11:53:07 & 2022.74 & 27.21 & 26.78 & 16.8$\pm$0.8\\
{\it Chandra} ACIS-S & 27467  & 2022-10-04T06:14:58 & 2022.76 & 19.81 & 19.62 & 14.9$\pm$0.9\\
{\it Chandra} ACIS-S & 27468  & 2022-10-09T17:27:37 & 2022.78 & 14.88 & 14.58 & 18.0$\pm$1.1\\
\hline
{\it XMM-Newton} EPIC-pn & 0304050101 & 2005-06-30T07:21:20 & 2005.50 & 70.64 & 35.75 & 53.3$\pm$1.3 \\
\hline
\end{tabular}
\label{tab:xobs}
\end{center}
\end{table*}

The first detection of X-ray emission from R Aqr was made with the {\it Einstein Observatory} in 1979 \citep{Jura1984}, followed by observations from {\it EXOSAT} and {\it ROSAT} \citep[][]{Viotti1987,Hunsch1998}. However, it was not until the advent of {\it Chandra} that the extended emission from the jet and that from the central engine were finally resolved \citep{Kellogg2001}. These authors showed that the spectrum of the central source of R Aqr was described by a combination of soft X-ray emission ($E<$2.0 keV) and a contribution at higher energies with a peak at 6.4 keV attributed to the Fe K$\alpha$ fluorescent line. This type of X-ray spectra from symbiotic systems is classified as a $\beta/\delta$-type \citep{Luna2013}. In combination with subsequent {\it Chandra} observations obtained by the end of year 2003, \citet{Kellogg2007} demonstrated that the X-ray spectra of the jets from R Aqr were best fit by plasma temperatures of $\sim$0.12--0.15 keV ($\approx$1.5$\times10^{6}$~K), very likely the result of an adiabatic shock produced by the jet velocity of a few times 10$^{2}$~km~s$^{-1}$.
More recently, \citet{Toala2022} presented the analysis of archival {\it XMM-Newton} observations obtained during 2005 and discovered the presence of extended super soft X-ray emission ($kT\approx0.09$ keV). This super soft emission fills the large-scale structures observed by nebular images of R Aqr and its present even when modelling the emission from the X-ray-emitting clumps from the jets. \citet{Toala2022} suggested that the origin of the extended super soft X-ray emission is material produced by the ongoing fast jets that has expanded and cooled down.

{\it Chandra} observations since 2017 were carried out as part of a multi-mission, multi-wavelength campaign to study the evolution of R Aqr during periastron passage \citep[PI: M.~Karovska; e.g.,][]{Huang2023,Sacchi2024}. By analysing {\it Chandra} and {\it Swift} X-ray observations around the periastron passage, \citet{Sacchi2024} reported an increase in the soft X-ray flux while the hard emission showed a decrease in flux. Several scenarios are proposed to explain this behaviour, focusing on changes in the physical properties or geometry of the surrounding medium, possibly as a result of the periastron passage. \citet{Sacchi2024} adopted Gaussian components to fit the 6.4 keV Fe fluorescent line in addition to power law models to fit the intermediate 2.0--4.0 keV energy range. Since the X-ray emission from symbiotic systems are expected to have a shock origin \citep{Mukai2017}, a more suitable model is warranted.

In this paper we present a reanalysis of the available X-ray {\it Chandra} and {\it XMM-Newton} observations of R Aqr. We include the effects of reflection physic models developed by our team \citep[e.g.,][]{Toala2023,Toala2024,Toala2024_single}, which help us dissecting the contributions from the different components of the X-ray spectra. In addition, we also used publicly available optical spectra to produce an unprecedented view of the accretion process in such an iconic symbiotic system such as R Aqr. We note that given the relatively large dispersion in the determination of an distance to R Aqr (see above), we will adopt a somewhat conservative distance value of $d$=200$^{+60}_{-20}$ pc, where the error values correspond to the different distance estimated by other authors. Given that parallax distance measurements of symbiotic stars are affected by significant uncertainties due to their extension, colours and orbital motion, we discount the distance estimated by {\it Gaia} \citep{Lindford2019,Sion2019}.

This paper is organised as follows. In Section~\ref{sec:obs} we describe the observations used in this work and their preparation. Section~\ref{sec:results} describes the analysis and the results. A discussion is presented in Section~\ref{sec:discussion} and, finally, our summary and conclusions are presented in Section~\ref{sec:conclusions}.

\begin{figure*}
\begin{center}
\includegraphics[width=0.5\linewidth]{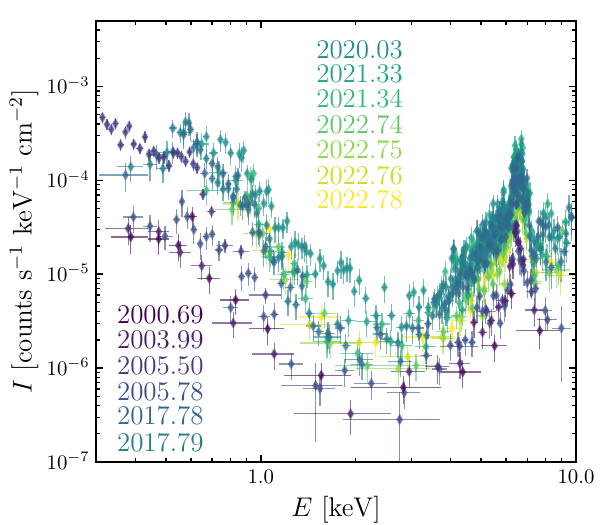}~
\includegraphics[width=0.5\linewidth]{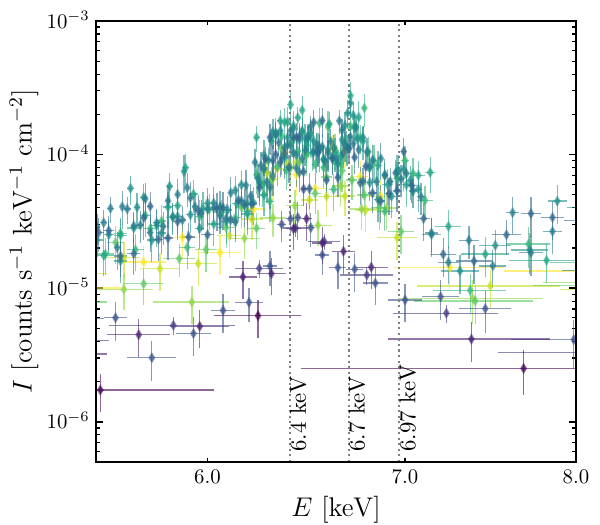}
\end{center}
\caption{Background-subtracted {\it Chandra} ACIS-S and {\it XMM-Newton} EPIC pn spectra of R~Aqr. Different epochs of observation are illustrated and labelled (see Table~\ref{tab:xobs} for details). The right panel shows a zoom into the 5.5--8.0 keV energy range to illustrate the contribution of the 6.4, 6.7 and 6.79 keV Fe emission lines.}
\label{fig:spec_all}
\end{figure*}

\section{Observations and Data preparation}
\label{sec:obs}

\subsection{X-ray data}

R Aqr is one of the most observed X-ray-emitting symbiotic systems. It has been observed by most major X-ray observatories. In this work we present a joint analysis of {\it Chandra} and {\it XMM-Newton} observations to study the impact of reflection physics on the estimation of the different spectral components of its X-ray emission. These observations are of particular interest because they cover the last 22~yr of evolution of R Aqr since its apastron in $\simeq$1998.7 and some years after periastron passage in $\simeq$2019.9 \citep{Sacchi2024}. 

R Aqr has been observed with the Advanced CCD Imaging Spectrometer (ACIS)-S on board {\it Chandra} at 12 different epochs between 2000 and 2022, corresponding to a total observing time of 446.75 ks. All observations were retrieved from the Chandra Data Archive\footnote{\url{https://cda.harvard.edu/chaser/}} and their details are listed in Table~\ref{tab:xobs}. The {\it Chandra} ACIS-S observations were processed with the Chandra Interactive Analysis of Observations \citep[{\sc ciao}, version 4.14;][]{Fruscione2006}. 

An event file per epoch was produced by reprocessing the data with the {\sc ciao} task \texttt{chandra\_repro}. After removing periods of high background, the observations did not reduce their net exposure times significantly (see Table~\ref{tab:xobs}).
The spectra of R Aqr were extracted from a circular aperture with 5~arcsec in radius with the {\sc ciao} task \texttt{specextract}. The background extraction region was defined by another circular aperture with 15~arcsec in radius in the vicinity of R Aqr with no contribution from this symbiotic system or the extended emission produced by the jets \citep[see][and references therein]{Sacchi2024}. The {\sc ciao} task \texttt{specextract} simultaneously produces the redistribution matrix and ancillary response files (RMF and ARF, respectively). All background-subtracted {\it Chandra} ACIS-S spectra of R Aqr are presented in Fig.~\ref{fig:spec_all}. Thre spectra were binned based on a minimum of 10 counts per bin. 

\begin{table*}
\begin{center}
\caption{Details of the optical spectra of R Aqr obtained from the ARAS spectral database. $F_\mathrm{opt}$ and $L_\mathrm{opt}$ are the estimated optical flux and luminosity of the accretion disc, respectively.  $\Delta \lambda$ denotes the complete wavelength range of each spectrum, while $\Delta \lambda_\mathrm{int}$ is the integration range used to estimate the optical flux ($F_\mathrm{opt}$) of the accretion disc. Here $\dot{M}_\mathrm{acc}$ has been computed including $L_\mathrm{UV}$ and $L_\mathrm{opt}$}.
\footnotesize
\setlength{\tabcolsep}{0.6\tabcolsep}  
\renewcommand{\arraystretch}{1.2}
\begin{tabular}{ccccccccccc}
\hline
Observation date & JD 2400000 & Observer  & Site  & $R$  & $\Delta \lambda$&  $\Delta \lambda_\mathrm{int}$ & M-type & $F_\mathrm{opt}$ & $L_\mathrm{opt}$ & $\dot{M}_\mathrm{acc}$ \\
        (UTC)&   (d)      &        &       &             &  (\AA)  & (\AA) & & (erg~cm$^{-2}$~s$^{-1}$) & (L$_\odot$) & (M$_\odot$~yr$^{-1}$)\\
\hline
2017-10-18T13:45 & 58045.073 & P. Luckas  & SPO-AU &   536 & 3700--7399 & 3860--6000 &M9 & $3.12\times 10^{-10}$& 0.39$^{+0.23}_{-0.08}$  &$(1.2^{+0.5}_{-0.3})\times10^{-9}$ \\
2019-11-15T10:32 & 58802.939 & T. Bohlsen & ARM-AU &  1760 & 3800--7400 & 3860--5450 &M6 & $6.26\times 10^{-11}$& 0.08$^{+0.05}_{-0.02}$  &$(4.5^{+1.9}_{-1.1})\times10^{-10}$\\
2020-01-12T02:02 & 58860.585 & F. Sims    & DCO-US &  1021 & 3715--7305 & 3860--5855 &M5 & $9.45\times 10^{-11}$& 0.12$^{+0.07}_{-0.02}$  &$(5.4^{+2.1}_{-1.4})\times10^{-10}$\\
2020-01-13T02:11 & 58861.591 & F. Sims    & DCO-US &  1021 & 3715--7305 & 3860--5597 &M5 & $9.03\times 10^{-11}$& 0.11$^{+0.07}_{-0.02}$  &$(5.3^{+2.0}_{-1.3})\times10^{-10}$\\
2020-01-14T02:12 & 58862.592 & F. Sims    & DCO-US &  1010 & 3716--7305 & 3860--5737 &M5 & $8.87\times 10^{-11}$& 0.11$^{+0.07}_{-0.02}$  &$(5.2^{+2.0}_{-1.3})\times10^{-10}$\\
2020-01-19T01:59 & 58867.583 & F. Sims    & DCO-US &  1066 & 3719--7305 & 3860--5738 &M6 & $7.22\times 10^{-11}$& 0.09$^{+0.05}_{-0.02}$  &$(4.8^{+1.8}_{-1.2})\times10^{-10}$\\
2020-07-30T08:51 & 59060.869 & F. Sims    & DCO-US &  1066 & 3719--7305 & 3860--5450 &M6 & $2.76\times 10^{-11}$& 0.35$^{+0.21}_{-0.07}$  &$(1.1^{+0.4}_{-0.3})\times10^{-9}$\\
2020-10-13T06:43 & 59135.780 & F. Sims    & DCO-US &   757 & 3710--7296 & 3860--6000 &M8 & $1.62\times 10^{-10}$& 0.20$^{+0.12}_{-0.04}$  &$(7.3^{+2.8}_{-2.0})\times10^{-10}$\\
2020-10-22T05:08 & 59144.714 & F. Sims    & DCO-US &  1057 & 3716--7304 & 3860--5930 &M8 & $1.48\times 10^{-10}$& 0.19$^{+0.11}_{-0.04}$  &$(6.9^{+2.7}_{-1.8})\times10^{-10}$\\
2020-11-26T03:07 & 59179.630 & F. Sims    & DCO-US &  1057 & 3716--7304 & 3860--5230 &M8 & $1.10\times 10^{-10}$& 0.14$^{+0.08}_{-0.03}$  &$(5.8^{+2.2}_{-1.5})\times10^{-10}$\\
2021-11-11T02:54 & 59529.621 & F. Sims    & DCO-US &  1032 & 3695--7283 & 3860--5700 &M8 & $1.29\times 10^{-10}$& 0.16$^{+0.10}_{-0.03}$  &$(6.3^{+2.4}_{-1.7})\times10^{-10}$\\
2022-10-02T06:40 & 59854.778 & F. Sims    & DCO-US &  1067 & 3718--7307 & 3860--6000 &M9 & $4.51\times 10^{-10}$& 0.56$^{+0.34}_{-0.11}$  &$(1.6^{+0.6}_{-0.5})\times10^{-9}$\\
2022-10-23T05:36 & 59875.734 & F. Sims    & DCO-US &  1026 & 3718--7307 & 3860--6000 &M9 & $3.00\times 10^{-10}$ & 0.38$^{+0.23}_{-0.07}$ &$(1.1^{+0.5}_{-0.3})\times10^{-9}$\\
2022-10-29T05:29 & 59881.729 & F. Sims    & DCO-US &  1069 & 3729--7291 & 3860--6000 &M9 & $2.71\times 10^{-10}$& 0.34$^{+0.20}_{-0.07}$  &$(1.0^{+0.4}_{-0.3})\times10^{-9}$\\
2022-11-01T04:43 & 59884.697 & F. Sims    & DCO-US &  1070 & 3728--7290 & 3860--6000 &M9 & $2.56\times 10^{-10}$& 0.32$^{+0.19}_{-0.06}$  &$(1.0^{+0.4}_{-0.3})\times10^{-9}$\\
2022-11-09T03:34 & 59892.649 & F. Sims    & DCO-US &  1066 & 3729--7300 & 3860--6000 &M9 & $2.21\times 10^{-10}$& 0.27$^{+0.16}_{-0.06}$  &$(9.0^{+3.6}_{-2.5})\times10^{-10}$\\
2022-11-17T03:58 & 59900.665 & F. Sims    & DCO-US &  1044 & 3729--730  & 3860--6000 &M9 & $1.82\times 10^{-10}$ & 0.23$^{+0.14}_{-0.05}$ &$(7.3^{+2.9}_{-2.0})\times10^{-10}$\\
2022-11-20T02:58 & 59903.624 & F. Sims    & DCO-US &  1084 & 3729--7300 & 3860--6000 &M9 & $1.62\times 10^{-10}$ & 0.20$^{+0.12}_{-0.04}$ &$(7.9^{+3.1}_{-2.1})\times10^{-10}$\\
2022-11-25T04:39 & 59908.694 & F. Sims    & DCO-US &  1048 & 3730--7301 & 3860--6000 &M9 & $1.76\times 10^{-10}$& 0.22$^{+0.13}_{-0.04}$  &$(7.7^{+3.0}_{-2.1})\times10^{-10}$\\
2022-11-30T04:09 & 59913.673 & F. Sims    & DCO-US &  1048 & 3727--7300 & 3860--6000 &M9 & $1.83\times 10^{-10}$ & 0.23$^{+0.14}_{-0.05}$ &$(7.9^{+3.1}_{-2.1})\times10^{-10}$\\
2022-12-10T03:28 & 59923.645 & F. Sims    & DCO-US &  1055 & 3730--7301 & 3860--6000 &M9 & $1.57\times 10^{-10}$&  0.20$^{+0.12}_{-0.04}$ &$(7.2^{+2.8}_{-1.9})\times10^{-10}$\\
\hline
\end{tabular}
\label{tab:opt_obs}
\end{center}
\end{table*}

\begin{figure*}
\includegraphics[width=0.75\linewidth]{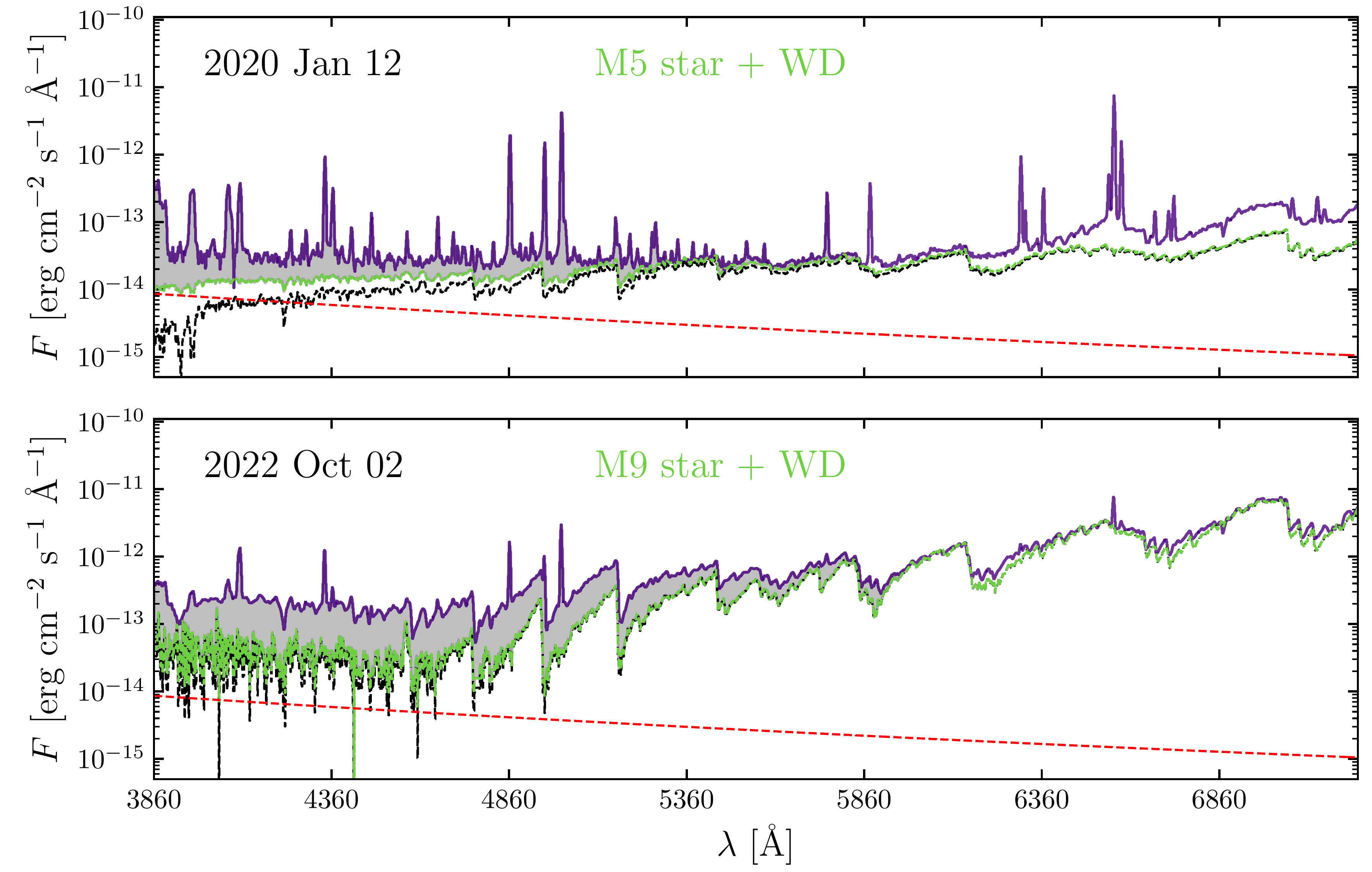}
\caption{Examples of extinction-corrected, flux-calibrated optical spectra of R Aqr (solid purple lines), obtained from the ARAS Spectral Database on 2020 Jan 12 (top) and 2022 Oct 02 (bottom). The spectra are compared with M-type star models from \citet{Fluks1994} (black dashed lines) and the contribution from the WD with the parameters estimated in Appendix~\ref{app:disc} ($T_\mathrm{eff}=25,000$ K and $L$=0.19~L$_\odot$; red dashed line). A combined spectrum (M-type star + WD) is shown with a (green) dashed line. The gray shaded area represents the contribution of the accretion disc to the optical flux.}

\label{fig:spec_opt}
\end{figure*}

R Aqr has been observed once by {\it XMM-Newton} using the European Photon Imaging Camera (EPIC) on 2005 Jun 30 (Obs. ID. 0304050101) with a total exposure time of 72.52 ks. The observation data files were retrieved from the XMM-Newton Science Archive\footnote{\url{https://www.cosmos.esa.int/web/xmm-newton/xsa}} and were processed following standard routines with the Science Analysis
Software \citep[{\sc sas}, version 20.0;][]{Gabriel2004} with the calibration matrices obtained on 2024 Jun 28. 

The EPIC spectra of R Aqr were extracted from a circular aperture with 1~arcmin in radius using the {\sc sas} task \texttt{evselect}. The background was defined by a similar aperture but from a region with no contribution from the extended X-ray emission originally reported in \citet{Toala2022}. The corresponding RMF and ARF files were produced with the \texttt{rmfgen} and \texttt{arfgen} tasks, respectively. Given its superior effective area with respect to the MOS detectors, in this paper we will only concentrate in the analysis of the EPIC pn spectrum. This is presented in Fig.~\ref{fig:spec_all} alongside the {\it Chandra} ACIS-S spectra of R Aqr.  
This EPIC pn spectrum is required to have a minimum of 60 counts per bin. 
The details of the {\it XMM-Newton} EPIC pn observations are also presented in Table~\ref{tab:xobs}.

\subsection{Optical data}
\label{sec:optical_data}

To complement the X-ray observations, we searched for publicly available contemporary optical data. Optical spectra of R Aqr were retrieved from the ARAS spectral database\footnote{\url{https://aras-database.github.io/database/index.html}} \citep{Teyssier2019}, which provides spectra from multiple epochs thanks to its international monitoring program of symbiotic stars. Only flux-calibrated spectra were retrieved from the ARAS database.
Table~\ref{tab:opt_obs} lists the details of the individual spectra analysed in this work (observer, observation location, spectral resolution, and wavelength range covered by each spectrum). These include spectra obtained between 2017 Oct 18 and 2022 Dec 10. 

The flux-calibrated ARAS spectra were inspected in {\sc iraf} \citep[][]{Tody1993} to calculate individual reddening $c$(H$\beta$) values for each spectrum from the H$\alpha$/H$\beta$ line ratio. However, we obtained H$\alpha$/H$\beta \gtrsim 3$ that might suggest that the recombination Case B might not be applicable to R Aqr. This issue is known for S-type symbiotic systems where the H$\alpha$/H$\beta$ ratio has been found to be reddening-free \citep{Mikolajewska1992,Mikolajewska1997}. Consequently, we decided to deredden the optical spectra by adopting the hydrogen column density ($N_\mathrm{H}$) reported towards the direction of R Aqr. The $N_\mathrm{H}$ column density tool of {\sc heasarc}\footnote{\url{https://heasarc.gsfc.nasa.gov/cgi-bin/Tools/w3nh/w3nh.pl}} suggests a value of 1.83$\times10^{20}$~cm$^{-2}$, which adopting $R_\mathrm{V}$=3.1 can be converted into an extinction $A_\mathrm{V}$ of 0.1.

The spectra were then corrected for extinction using the \texttt{remove} module from the \texttt{extinction}\footnote{\url{https://extinction.readthedocs.io/en/latest/index.html}} Python library. The extinction function is calculated using the \texttt{ccm89} module from the same library, which is based on the extinction law by \citet{Cardelli1998}. In Fig.~\ref{fig:spec_opt} we present two examples of flux-calibrated, extinction-corrected ARAS spectra that are contemporaneous with two of the X-ray epochs of observations, while the rest of them are presented in Appendix~\ref{app:aras}.

\section{Analysis and results} 
\label{sec:results}

The X-ray spectra presented in Fig.~\ref{fig:spec_all} illustrate the X-ray evolution of R Aqr over a time span of 22 yr from 2000 Oct to 2022 Sep, covering about half of the orbital period of the system. 
The left panel of Fig.~\ref{fig:spec_all} shows noticeable variations in the total apparent X-ray flux of R\,Aqr in the 0.3--10.0 keV. 
The earliest observations (2000.69 and 2003.99) have the faintest observed fluxes, while there is a peak over epochs $\sim$2017--2020, followed by an apparent decrease to the last observation during 2022. 

The right panel of Fig.~\ref{fig:spec_all} shows the variation of the three Fe emission lines for the 22 yr period studied here.  
The Fe fluorescent line at 6.4 keV is present in all epochs, thus suggesting that reflection is an unavoidable component. 
At early epochs, the 5.5--8.0 keV energy range is dominated by the 6.4 keV fluorescent line, with the contribution from the Fe emission lines at 6.7 and 6.97 keV becoming relevant at epochs later than 2005. 
In fact, the flux in this energy range increases with time suggesting that the flux of the heavily-extinguished plasma component associated with the boundary layer increased its flux in the later epochs.

The analysis of the X-ray spectra was performed with the X-Ray Spectral Fitting Package {\sc xspec} \citep[version 12.12.1;][]{Arnaud1996}. Following previous analyses of the X-ray observations of R Aqr \citep{Kellogg2001,Kellogg2007,Sacchi2024} and other $\beta/\delta$-type symbiotic stars \citep[e.g.,][]{Karovska2007,Zhekov2019}, we adopted two distinct hydrogen column density components ($N_\mathrm{H1}$ and $N_\mathrm{H2}$) using the Tuebingen-Boulder ISM absorption model \citep[\texttt{tbabs};][]{Wilms2000} included in {\sc xspec}. $N_\mathrm{H1}$ is usually found to affect the soft spectral region very likely produced by the extended emission, while the second and larger extinction $N_\mathrm{H2}$ affects the emission from the boundary layer and the contribution from the reflection component. For the X-ray emission from R Aqr, we adopted an emission spectrum from collisionally-ionised diffuse gas, specifically, the \texttt{apec} model included in {\sc xspec}\footnote{\url{https://heasarc.gsfc.nasa.gov/xanadu/xspec/manual/node134.html}}. The abundances were set to Solar values from \citet{Lodders2009}.

\begin{table}
\begin{center}
\caption{Details of the flared disc model obtained with {\sc skirt}. See Fig.~\ref{fig:disc_fig} for further details.} 
\begin{tabular}{ll}
\hline
Parameter & Value  \\
\hline
$r_\mathrm{in}$&5.4$\times10^{-4}$~AU\\
$r_\mathrm{out}$ &5~AU\\
$\phi$&30$^{\circ}$ \\
$i$  & 20$^{\circ}$\\
$N_\mathrm{H,ref}$&  5$\times10^{24}$ cm$^{-2}$\\
$T$ & 10$^{4}$~K\\
\hline
\end{tabular}
\label{tab:param}
\end{center}
\end{table}

\subsection{The accretion disc as a source of reflection}

The ubiquitous presence of the fluorescent Fe emission line at 6.4 keV in the X-ray spectra of R Aqr suggests a more complex and physically-meaningful reflection-based analysis of the emission than the often used Gaussian profile. Indeed, reflection from the accretion disc around the WD component in symbiotic systems has been used before to study the impact on the presence of the 6.4 keV Fe emission line \citep[e.g.,][]{Ishida2009,LdO2018,Toala2023,Toala2024} and a similar scenario will be adopted here for R Aqr.

Using the best estimates for the stellar components and their orbital parameters, we can approximate the size of the accretion disc. Recent estimates of the semi-major axis of the binary orbit predict $a$=47--57~mas \citep{Alcolea2023,Bujarrabal2018} which, at a distance of $d$=200$^{+60}_{-20}$ pc, translates to a maximum value of $a=5.5^{+1.7}_{-0.6}\times10^{-5}$~pc or $a$=11.4$^{+3.5}_{-1.1}$~AU. 

Using the approximation for the effective Roche lobe radius \citep[see][]{Eggleton1983}
\begin{equation}
    \frac{r_\mathrm{L}}{a} = \frac{0.49 q^{2/3}}{0.6q^{2/3} + \ln(1 + q^{1/3})},
\end{equation}
\noindent where $q$ is the companion to the WD mass ratio, we obtain $r_\mathrm{L}$=4.7$^{+1.4}_{-0.5}$~AU for a WD mass of $M_\mathrm{WD}$=0.7 M$_\odot$ and a mass for the red giant companion of 1.0~M$_\odot$ \citep[following the mass ratio of 1.4 and a total mass of the binary of 1.7$\pm$0.1~M$_\odot$ reported in][]{Alcolea2023}. We note that \citet{Hinkle2022} proposed a very similar radius of 5~AU for the accretion disc.

We have generated reflection models of disc-like geometries using the {\sc skirt} radiative transfer code \citep[version 9;][]{Camps2020} that currently includes the calculations for the radiative transfer of X-ray photons \citep{VanderMeulen2023}. We fixed the geometry of the disc to a flared disc with an opening angle of $\phi=30^\circ$ and inner and outer radii of $r_\mathrm{in}$=5.4$\times10^{-4}$~AU (=0.012~R$_\odot$)\footnote{Typical radius of WDs are estimated to be about 6000--8000~km. Here we adopt the later value for $r_\mathrm{in}$.} and $r_\mathrm{out}$=5~AU, respectively\footnote{ We note that \citet{Paczynski1977} suggested that $r_\mathrm{out} \leq r_\mathrm{L}$ and the selection of $r_\mathrm{out}=5$ AU in fact agrees with this relationship within the error values estimated for $r_\mathrm{L}$.}. Several values for the averaged column density of the disc were explored ($N_\mathrm{H,ref}$=10$^{24}$, 5$\times10^{24}$, 10$^{25}$, 5$\times10^{25}$, and 10$^{26}$~cm$^{-2}$), but the best quality fits to the X-ray data resulted when using $N_\mathrm{H,ref}$=5$\times10^{24}$~cm$^{-2}$. 
Finally, the accretion disc is assumed to be ionised with a temperature of $T$=10$^{4}$~K.

\begin{figure}
\includegraphics[width=\linewidth]{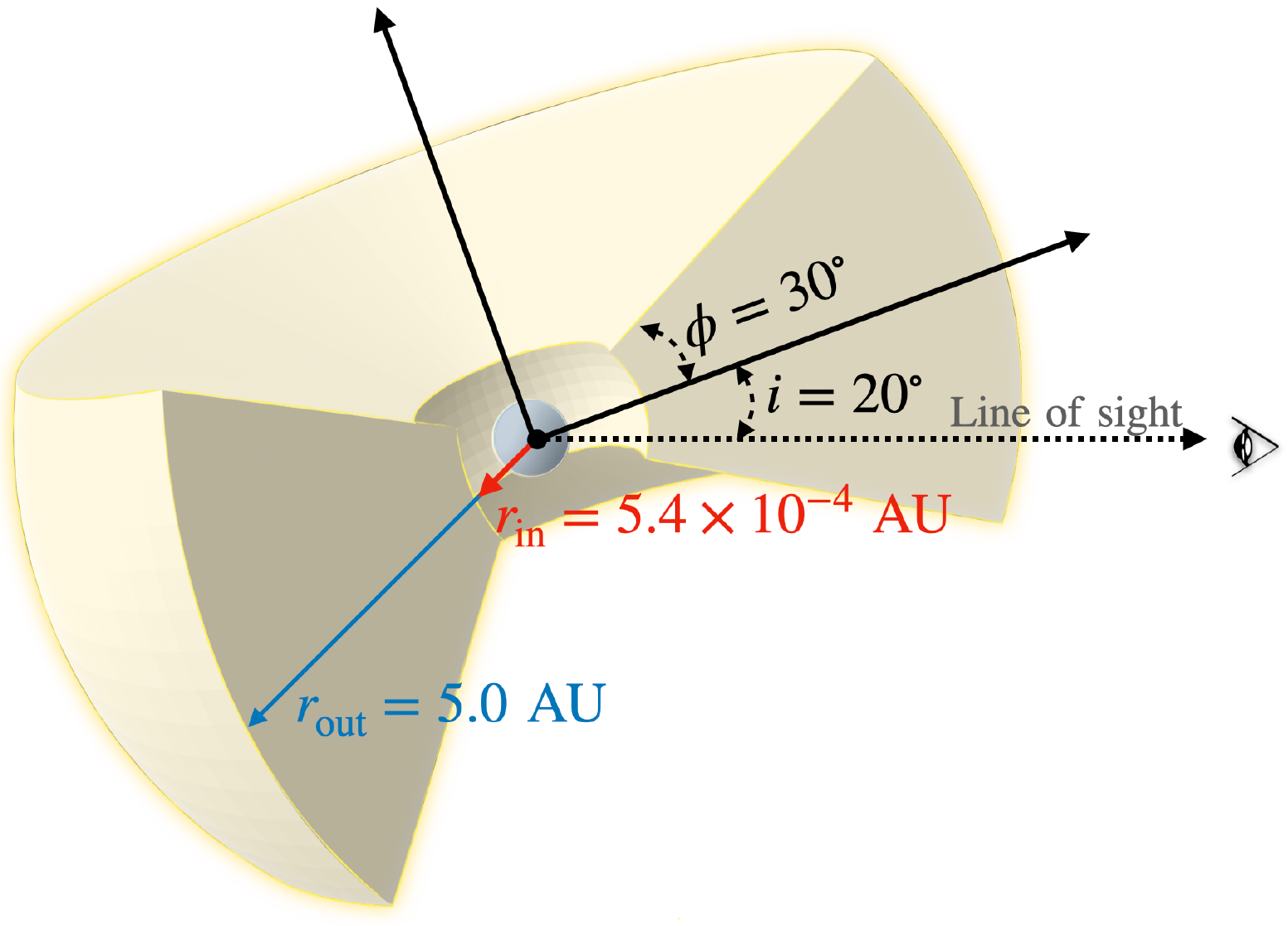}
\caption{Representation of the accretion disc model around the WD component used in our {\sc skirt} radiative transfer models. The diagram shows its orientation with respect to the line of sight.}
\label{fig:disc_fig}
\end{figure}

\begin{table*}
\caption{Parameters of the best fit models of the EPIC-pn observations of R Aqr. The fluxes were computed in the 0.3--10.0 keV and are presented in cgs units (erg~cm$^{-2}$~s$^{-1}$). The adopted distance is $d$=200 pc. Boldface values represent fixed values. The normalisation parameter is defined in {\sc xspec} as $A\approx \int n_\mathrm{e}^2 dV/4 \pi d^2$, where $n_\mathrm{e}$ is the electron number density. Here $L_\mathrm{acc}$ and $\dot{M}_\mathrm{acc}$ are computed including $L_\mathrm{X3}$, $L_\mathrm{UV}$ and $L_\mathrm{opt}$} (see Section~\ref{sec:discussion} for details).
\setlength{\tabcolsep}{0.4\tabcolsep}  
\scriptsize
\label{tab:obs2}
\begin{center}
\begin{tabular}{ccccccccc}
\hline
Parameter & Units & 2000.69 & 2003.99 & 2005.50 & 2005.78  & 2017.78 & 2017.79 & 2020.03 \\
\hline
$\chi^{2}_\mathrm{DoF}$   &  & 5.33/5=1.07  & 14.20/13=1.09 & 41.50/28=1.48 & 57.76/44=1.31 & 142.95/108=1.32 & 221.45/183=1.21 &  158.42/148=1.07 \\
\\ 
$N_\mathrm{H1}$ & [10$^{21}$~cm$^{-2}$] & 5.3$\pm$2.6 & 3.73$\pm$0.4 & 1.48$\pm$0.70 & 4.8$\pm$1.1 & 4.7$\pm$0.8 & 4.7$\pm$0.1 & 2.6$\pm$1.0\\
$kT_1$ & [keV] & {\bf 0.025} & 0.033$\pm$0.001 & 0.040$\pm$0.016 & 0.020$\pm$0.001 & 0.042$\pm$0.011 & 0.047$\pm$0.008 & 0.038$\pm$0.001\\
$A_1$  & [cm$^{-5}$] & 390$\pm$150 & 3.2$\pm$1.1 & 0.32$\pm$0.02 & 14900$\pm$300 & 8.9$\pm$1.9 & 10.5$\pm$1.5 & 10.0$\pm$2.9\\
$f_\mathrm{X1}$  & [cgs] & (2.5$\pm$0.9)$\times10^{-15}$  & (4.9$\pm$1.6)$\times10^{-15}$ & (9.0$\pm$0.5)$\times10^{-14}$ & (3.6$\pm$0.3)$\times10^{-15}$ & (6.6$\pm$1.1)$\times10^{-14}$ & (7.7$\pm$0.9)$\times10^{-14}$ & (2.6$\pm$0.7)$\times10^{-13}$ \\
$F_\mathrm{X1}$  & [cgs] & (3.0$\pm$1.1)$\times10^{-11}$  & (5.0$\pm$1.8)$\times10^{-12}$ & (3.4$\pm$0.1)$\times10^{-12}$ & (8.7$\pm$0.2)$\times10^{-11}$ & (8.5$\pm$1.8)$\times10^{-11}$ & (1.0$\pm$0.1)$\times10^{-10}$ & (4.9$\pm$1.4)$\times10^{-11}$\\
$kT_2$ & [keV] & 0.13$\pm$0.04 & 0.17$\pm$0.04 & 0.17$\pm$0.01 & 0.13$\pm$0.03 & 0.20$\pm$0.03 & 0.20$\pm$0.02 & 0.28$\pm$0.04\\
$A_2$  & [cm$^{-5}$] & (2.7$\pm$1.9)$\times10^{-4}$ & (3.0$\pm$2.4)$\times10^{-5}$ & (8.1$\pm$0.8)$\times10^{-5}$ & (5.9$\pm$1.8)$\times10^{-4}$ & (6.0$\pm$4.4)$\times10^{-4}$ & (7.0$\pm$3.6)$\times10^{-4}$ & (2.6$\pm$1.6)$\times10^{-4}$ \\
$f_\mathrm{X2}$  & [cgs] & (5.0$\pm$3.5)$\times10^{-15}$ & (2.8$\pm$2.2)$\times10^{-15}$ & (3.6$\pm$0.4)$\times10^{-14}$ & (1.2$\pm$0.3)$\times10^{-14}$ & (5.6$\pm$3.9)$\times10^{-14}$ & (6.5$\pm$3.2)$\times10^{-14}$ & (1.3$\pm$0.8)$\times10^{-13}$ \\
$F_\mathrm{X2}$  & [cgs] & (2.2$\pm$1.5)$\times10^{-13}$ & (4.9$\pm$3.8)$\times10^{-14}$ & (1.4$\pm$0.1)$\times10^{-13}$ & (6.1$\pm$1.9)$\times10^{-13}$ & (1.2$\pm$0.8)$\times10^{-12}$ & (1.4$\pm$0.6)$\times10^{-12}$ & (5.7$\pm$3.7)$\times10^{-13}$ \\
\\
$N_\mathrm{H2}$ & [10$^{21}$~cm$^{-2}$] & 158.9$\pm$133.3 & 295.4$\pm$51.4 & 359.8$\pm$25.7 & 120.0$\pm$15.3 & 410$\pm$25.3 & 410.9$\pm$18.3 & 462.3$\pm$25.6\\
CF  & & 0.910$\pm$0.080 & 0.993$\pm$0.003 & 0.993$\pm$0.007 & 0.975$\pm$0.014 & 0.997$\pm$0.002 & 0.997$\pm$0.001 & 0.991$\pm$0.002\\
$kT_3$  & [keV] & {\bf 8.0} & 4.1$\pm$1.6 & 11.9$\pm$5.0 & 10.0$\pm$8.2 & 6.9$\pm$0.7 & 6.9$\pm$0.5 & 6.0$\pm$0.5\\
$A_{3}$ & [cm$^{-5}$] & (6.0$\pm$5.3)$\times10^{-5}$ & (3.8$\pm$3.7)$\times10^{-4}$ & (4.1$\pm$1.4)$\times10^{-4}$ & (8.6$\pm$3.5)$\times10^{-5}$ & (3.3$\pm$0.4)$\times10^{-3}$ & (3.9$\pm$0.3)$\times10^{-3}$ & (5.3$\pm$0.5)$\times10^{-3}$ \\
$f_\mathrm{X3}$ & [cgs] & (4.5$\pm$4.1)$\times10^{-14}$ & (1.1$\pm$1.0)$\times10^{-13}$ & (1.8$\pm$0.6)$\times10^{-13}$ & (6.8$\pm$3.0)$\times10^{-14}$ & (1.1$\pm$0.1)$\times10^{-12}$ & (1.2$\pm$0.2)$\times10^{-12}$ & (1.5$\pm$0.1)$\times10^{-12}$\\
$F_\mathrm{X3}$ & [cgs] & (1.2$\pm$1.1)$\times10^{-13}$ & (6.9$\pm$6.8)$\times10^{-13}$ & (8.7$\pm$3.1)$\times10^{-13}$ & (1.8$\pm$0.8)$\times10^{-13}$ & (6.8$\pm$0.9)$\times10^{-12}$ & (7.9$\pm$0.8)$\times10^{-12}$ & (1.1$\pm$0.1)$\times10^{-11}$\\
$A_\mathrm{ref}$ & [cm$^{-5}$] & (5.6$\pm$1.6)$\times10^{-3}$ & (1.8$\pm$0.4)$\times10^{-2}$ & (2.0$\pm$0.3)$\times10^{-2}$ & (1.8$\pm$0.2)$\times10^{-2}$ & (6.2$\pm$0.6)$\times10^{-2}$ & (7.2$\pm$0.5)$\times10^{-2}$ & (8.8$\pm$0.8)$\times10^{-2}$ \\
$f_\mathrm{ref}$ & [cgs] & (8.6$\pm$2.5)$\times10^{-14}$ & (2.1$\pm$0.4)$\times10^{-13}$ & (2.1$\pm$0.3)$\times10^{-13}$ & (2.8$\pm$0.3)$\times10^{-13}$ & (5.7$\pm$0.6)$\times10^{-13}$ & (6.7$\pm$0.4)$\times10^{-13}$ & (7.5$\pm$0.7)$\times10^{-13}$ \\
$F_\mathrm{ref}$ & [cgs] & (1.1$\pm$0.3)$\times10^{-13}$ & (3.6$\pm$0.6)$\times10^{-13}$ & (3.9$\pm$0.6)$\times10^{-13}$ & (3.5$\pm$0.4)$\times10^{-13}$ & (1.2$\pm$0.1)$\times10^{-12}$ & (1.4$\pm$0.1)$\times10^{-12}$ & (1.7$\pm$0.2)$\times10^{-12}$\\
\\
$f_\mathrm{X,TOT}$ & [cgs] & (1.4$\pm$0.7)$\times10^{-13}$ & (3.2$\pm$1.5)$\times10^{-13}$ & (5.1$\pm$)$\times10^{-13}$ & (3.7$\pm$0.6)$\times10^{-13}$ & (1.8$\pm$0.2)$\times10^{-12}$ & (2.1$\pm$0.5)$\times10^{-12}$ & (2.6$\pm$0.4)$\times10^{-12}$ \\
$F_\mathrm{X,TOT}$ & [cgs] & (3.0$\pm$1.2)$\times10^{-11}$ & (6.2$\pm$2.5)$\times10^{-12}$ & (4.8$\pm$0.5)$\times10^{-12}$ & (8.9$\pm$0.2)$\times10^{-11}$ & (9.4$\pm$1.9)$\times10^{-11}$ & (1.1$\pm$0.2)$\times10^{-10}$ & (6.2$\pm$1.5)$\times10^{-11}$ \\
$L_\mathrm{X,TOT}$ & [erg~s$^{-1}$] & (1.4$\pm$0.6)$\times10^{32}$ & (3.0$\pm$1.1)$\times10^{31}$ & (2.3$\pm$0.2)$\times10^{31}$ & (4.2$\pm$0.1)$\times10^{32}$ & (4.5$\pm$0.9)$\times10^{32}$ & (5.2$\pm$0.8)$\times10^{32}$ & (2.9$\pm$0.7)$\times10^{32}$ \\
\\
$F_\mathrm{opt}$ & [erg~cm$^{-2}$~s$^{-1}$] & \dots & \dots & \dots & \dots & (2.90$\pm$0.72)$\times10^{-10}$ & (2.89.$\pm$0.72)$\times10^{-10}$ & (1.57$\pm$0.30)$\times10^{-10}$\\
$L_\mathrm{opt}$ & [L$_\odot$] & \dots & \dots & \dots & \dots & $0.36^{+0.14}_{-0.10}$ &$0.36^{+0.14}_{-0.10}$&$0.20^{+0.07}_{-0.04}$\\
$L_\mathrm{acc}$ & [L$_\odot$] & \dots & \dots & \dots & \dots & $0.44^{+0.14}_{-0.10}$ &$0.44^{+0.14}_{-0.11}$&$0.28^{+0.07}_{-0.04}$\\
$\dot{M}_\mathrm{acc}$ & [M$_\odot$ yr$^{-1}$]& \dots & \dots & \dots & \dots & $(9.9^{+4.4}_{-2.4})\times10^{-10}$ & $(9.9^{+4.4}_{-2.4})\times10^{-10}$ & $(6.3^{+2.5}_{-1.2})\times10^{-10}$ \\
$\eta$ & [10$^{-3}$]& \dots & \dots & \dots & \dots  & 9.9$^{+4.4}_{-2.4}$ & 9.9$^{+4.4}_{-2.4}$ & 6.3$^{+2.5}_{-1.2}$ \\
\hline
Parameter & Units & 2021.33 & 2021.34 & 2022.74 & 2022.75 & 2022.76 & 2022.78 \\
\hline
$\chi^{2}_\mathrm{DoF}$ &  & 113.79/84=1.35 & 117.61/72=1.63 & 38.13/31=1.23 & 13.85/12=1.15 & 19.83/19=1.04 & 17.71/16=1.11\\ 
$N_\mathrm{H1}$ & [10$^{21}$~cm$^{-2}$] & 1.0$\pm$1.8 & 5.5$\pm$1.5 & 7.5$\pm$3.9 & 6.9$\pm$4.8 & 4.6$\pm$12.4 & 2.5$\pm$14.5\\
$kT_2$ & [keV] & 0.21$\pm$0.06 & 0.19$\pm$0.05 & 0.15$\pm$0.08 & 0.12$\pm$0.03 & 0.12$\pm$0.11 & 0.19$\pm$0.06\\
$A_2$  & [cm$^{-5}$] & (1.7$\pm$0.2)$\times10^{-4}$ & (1.2$\pm$1.7)$\times10^{-3}$ & (6.2$\pm$1.0)$\times10^{-3}$ & (1.8$\pm$0.5)$\times10^{-2}$ & (9.5$\pm$2.2)$\times10^{-3}$ & (5.3$\pm$1.4)$\times10^{-4}$ \\
$f_\mathrm{X2}$  & [cgs] & (1.6$\pm$0.1)$\times10^{-13}$ & (6.5$\pm$9.5)$\times10^{-14}$ & (5.2$\pm$0.9)$\times10^{-14}$ & (4.6$\pm$2.2)$\times10^{-14}$ & (1.3$\pm$0.3)$\times10^{-13}$ & (9.7$\pm$3.8)$\times10^{-14}$\\
$F_\mathrm{X2}$  & [cgs] & (3.5$\pm$0.4)$\times10^{-13}$ & (2.2$\pm$3.2)$\times10^{-12}$ & (8.3$\pm$1.4)$\times10^{-12}$ & (1.3$\pm$0.5)$\times10^{-11}$ & (7.5$\pm$1.8)$\times10^{-12}$ & (9.0$\pm$3.4)$\times10^{-13}$ \\
\\
$N_\mathrm{H2}$ & [10$^{21}$~cm$^{-2}$] & 511.8$\pm$43.7 & 332.2$\pm$31.4 & 473.0$\pm$57.2 & 468.6$\pm$9.0 & 444.6$\pm$69.2 & 551.6$\pm$87.6 \\
CF  & & 0.994$\pm$0.002 & 0.994$\pm$0.003 & 0.997$\pm$0.003 & 0.992$\pm$0.005 & 0.994$\pm$0.004 & 0.992$\pm$0.004\\
$kT_3$  & [keV] & 6.4$\pm$1.0 & 5.9$\pm$1.1 & 6.0$\pm$1.3 & 5.5$\pm$2.0 & 6.7$\pm$1.7 & 4.7$\pm$1.8 \\
$A_{3}$ & [cm$^{-5}$] & (3.9$\pm$0.7)$\times10^{-3}$ & (2.1$\pm$0.4)$\times10^{-3}$ & (2.1$\pm$0.5)$\times10^{-3}$ & (1.5$\pm$0.6)$\times10^{-3}$ & (1.7$\pm$0.5)$\times10^{-3}$ & (3.2$\pm$1.2)$\times10^{-3}$\\
$f_\mathrm{X3}$ & [cgs] & (1.0$\pm$0.2)$\times10^{-12}$ & (7.6$\pm$1.3)$\times10^{-13}$ & (5.6$\pm$1.2)$\times10^{-13}$ & (3.9$\pm$1.5)$\times10^{-13}$ & (5.0$\pm$1.6)$\times10^{-13}$ & (6.0$\pm$2.4)$\times10^{-13}$ \\
$F_\mathrm{X3}$ & [cgs] & (8.0$\pm$1.4)$\times10^{-12}$ & (4.3$\pm$0.7)$\times10^{-12}$ & (4.3$\pm$1.0)$\times10^{-12}$ & (3.0$\pm$1.2)$\times10^{-12}$ & (3.4$\pm$1.1)$\times10^{-12}$ & (6.0$\pm$2.4)$\times10^{-12}$ \\
$A_\mathrm{ref}$ & [cm$^{-5}$] & 0.15$\pm$0.01  & (7.0$\pm$0.6)$\times10^{-2}$ & (6.4$\pm$0.9)$\times10^{-2}$ & (3.5$\pm$0.8)$\times10^{-2}$ & (5.7$\pm$0.9)$\times10^{-2}$ & (7.7$\pm$1.5)$\times10^{-2}$ \\
$f_\mathrm{ref}$ & [cgs] & (1.2$\pm$0.1)$\times10^{-12}$ & (7.5$\pm$0.6)$\times10^{-13}$ & (5.3$\pm$0.8)$\times10^{-13}$ & (2.9$\pm$0.7)$\times10^{-13}$ & (5.0$\pm$0.8)$\times10^{-13}$ & (5.6$\pm$1.2)$\times10^{-13}$ \\
$F_\mathrm{ref}$ & [cgs] & (2.9$\pm$0.4)$\times10^{-12}$ & (1.4$\pm$0.1)$\times10^{-12}$ & (1.3$\pm$0.1)$\times10^{-12}$ & (6.7$\pm$1.7)$\times10^{-13}$ & (1.1$\pm$0.2)$\times10^{-12}$ & (1.5$\pm$0.3)$\times10^{-12}$ \\
\\
$f_\mathrm{X,TOT}$ & [cgs] & (2.4$\pm$0.3)$\times10^{-12}$ & (1.6$\pm$0.3)$\times10^{-12}$ & (1.1$\pm$0.3)$\times10^{-12}$ & (7.3$\pm$2.3)$\times10^{-13}$ & (1.1$\pm$0.3)$\times10^{-12}$ & (1.3$\pm$0.3)$\times10^{-12}$ \\
$F_\mathrm{X,TOT}$ & [cgs] & (1.1$\pm$0.2)$\times10^{-11}$ & (7.8$\pm$4.0)$\times10^{-12}$ & (1.4$\pm$0.2)$\times10^{-11}$ & (1.7$\pm$0.6)$\times10^{-11}$ & (1.2$\pm$0.3)$\times10^{-11}$ & (8.4$\pm$3.0)$\times10^{-12}$\\
$L_\mathrm{X,TOT}$ & [erg~s$^{-1}$] & (5.2$\pm$0.9)$\times10^{31}$ & (3.7$\pm$1.9)$\times10^{31}$ & (6.7$\pm$1.1)$\times10^{31}$ & (8.1$\pm$2.7)$\times10^{31}$ & (5.7$\pm$1.5)$\times10^{31}$ & (4.0$\pm$1.4)$\times10^{31}$\\
\\

$F_\mathrm{opt}$ & [erg~cm$^{-2}$~s$^{-1}$] & (2.11$\pm$0.36)$\times10^{-10}$ & (2.12$\pm$0.36)$\times10^{-10}$ & (2.72$\pm$0.24)$\times10^{-10}$ & (2.72$\pm$0.24)$\times10^{-10}$ & (2.72$\pm$0.24)$\times10^{-10}$ & (2.72$\pm$0.24)$\times10^{-10}$\\
$L_\mathrm{opt}$ & [L$_\odot$] & $0.26^{+0.01}_{-0.05}$ & $0.27^{+0.01}_{-0.05}$ & $0.34^{+0.11}_{-0.05}$ & $0.34^{+0.11}_{-0.05}$& $0.34^{+0.11}_{-0.05}$ & $0.34^{+0.11}_{-0.05}$\\
$L_\mathrm{acc}$ & [L$_\odot$] & $0.34^{+0.09}_{-0.05}$ & $0.34^{+0.09}_{-0.05}$ & $0.41^{+0.11}_{-0.05}$ &  $0.41^{+0.11}_{-0.05}$& $0.41^{+0.11}_{-0.05}$&  $0.41^{+0.11}_{-0.05}$\\
$\dot{M}_\mathrm{acc}$& [M$_\odot$ yr$^{-1}$]& $(7.7^{+3.1}_{-1.4})\times10^{-10}$ &  $(7.7^{+3.1}_{-1.4})\times10^{-10}$  & $(9.3^{+3.7}_{-1.4})\times10^{-10}$ &  $(9.3^{+3.7}_{-1.4})\times10^{-10}$ &  $(9.3^{+3.7}_{-1.4})\times10^{-10}$  &  $(9.4^{+3.7}_{-1.4})\times10^{-10}$   \\

$\eta$ & [10$^{-3}$]& 7.7$^{+3.1}_{-1.4}$ & 7.6$^{+3.1}_{-1.4}$ & 9.3$^{+3.7}_{-1.4}$ & 9.3$^{+3.7}_{-1.4}$ & 9.3$^{+3.7}_{-1.4}$ & 9.4$^{+3.7}_{-1.4}$ \\
\hline
\end{tabular}
\end{center}
\end{table*}

\begin{figure*}
\begin{center}
\includegraphics[width=0.9\linewidth]{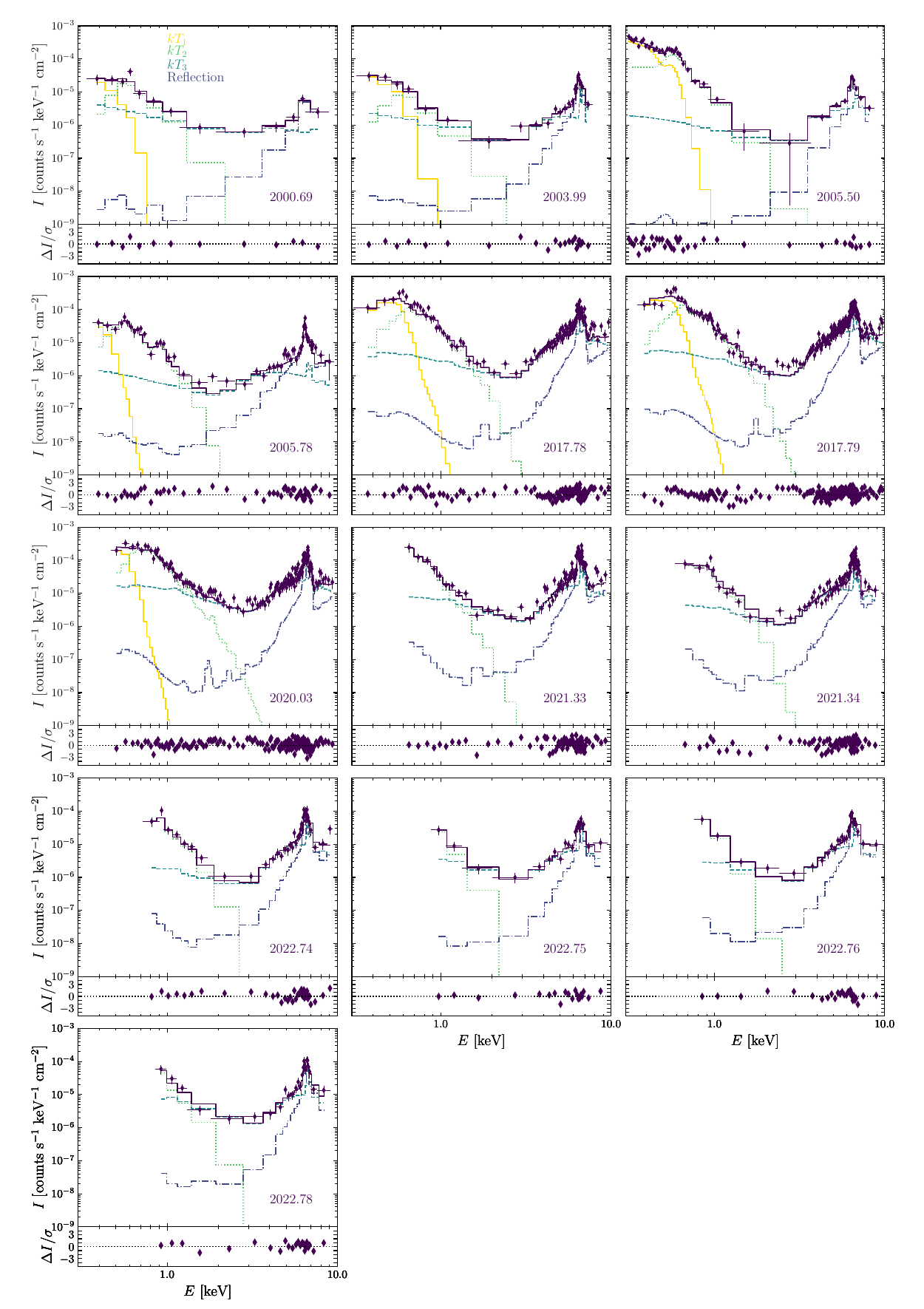}
\caption{Background-subtracted X-ray spectra of R Arq. Different panels show details of the best-fit models for each epoch. The contribution from the different components to each epoch are shown. See details in Table~\ref{tab:obs2}.}
\label{fig:xspec}
\end{center}
\end{figure*}

\citet{Bujarrabal2021} presented numerical simulations of the interaction of a WD accreting material from a mass-loosing AGB star to model the density distribution of the R Aqr system and discussed that the inclination angle between the orbital plane and the line of sight should be $i \lesssim 30^{\circ}$, and  \citet{Bujarrabal2021} and \citet{Alcolea2023} suggested that the best fit to the orbital parameters is $i$=20$^{\circ}$. Thus, we fixed the inclination to $i=20^{\circ}$. The reflection model obtained with {\sc skirt} was converted to an additive single-parameter table (the normalisation parameter $A_\mathrm{ref}$) using the {\sc heasoft}\footnote{\url{https://heasarc.gsfc.nasa.gov/docs/software/heasoft/}} task \texttt{ftflx2tab}. As a summary, we list in Table~\ref{tab:param} all the parameters of the disc structure to produce the reflection component which is further illustrated in Fig.~\ref{fig:disc_fig}.

\subsection{The best-fit models of the X-ray spectra}

We adopted models in {\sc xspec} of the form:
\begin{equation}
    {\rm tbabs}_1 \cdot ({\rm apec}_1 + \mathrm{apec_2}) + 
    \\
    {\rm CF} \cdot {\rm tbabs}_3 \cdot (\mathrm{apec}_3 + {\rm reflection}),
\end{equation}
\noindent where CF is a covering factor that helps alleviating the fact that the disc and the circumstellar medium around R Aqr are not homogeneous. 

We first tried a single simultaneous fit to all X-ray spectra and produced an acceptable fit, but individually fitting each epoch improves the fit statistics with reduced $\chi^{2}_\mathrm{DoF}$ values between 1.07 and 1.63. The details of the different models are presented in Table~\ref{tab:obs2}, which lists the observed ($f_\mathrm{X,TOT}$) and intrinsic ($F_\mathrm{X,TOT}$) total fluxes and the properties of the different components for each epoch. 
The models and their components are compared with the background-subtracted spectra in Fig.~\ref{fig:xspec}.

The 2021.33--2022.78 epochs did not require the presence of the super soft plasma component denoted as $kT_1$ in our model, while the rest of the earlier epochs required it to have values between $kT_1$=0.02 and 0.047 keV. 
We attribute this situation to the reduction of the effective area of the {\it Chandra} ACIS instruments at soft energies. To assess the lack of the super soft component in the later epochs, we extracted spectra of a background source located 3.93~arcmin towards the southwest from R Aqr, at ($\alpha,\delta)$=(23$^\mathrm{h}$:43$^\mathrm{m}$:35.96$^\mathrm{s}$, $-$15$^\circ$:19$'$:16.93$''$). 
We corroborated that this source has a significant soft flux ($E<1.0$ keV) at epochs earlier than 2021 that is not detected in more recent observations.

The second soft component (detected in all epochs) resulted in plasma temperatures of $kT_2\lesssim0.2$ keV, similar to the estimates of previous works \citep[see][and references therein]{Sacchi2024}. The temperature of the plasma in the boundary layer varied between $kT_3=4$ and 12 keV, with a relatively constant plasma temperature of $kT_3\approx6$ keV for the 2017.78--2022.78 epochs. 
We particularly note that the most problematic epoch was  2000.69 because the insufficient photon statistics. 
The fit to this epoch required fixing $kT_1$ and $kT_3$ at 0.025 and 8.0 keV, respectively.

\begin{figure}
\includegraphics[width=\linewidth]{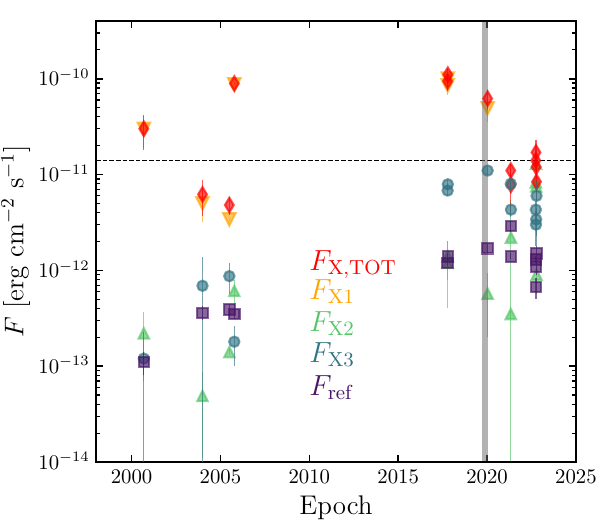}
\caption{Evolution with time of the intrinsic X-ray flux in R~Aqr. Different symbols show the evolution of the total flux ($F_\mathrm{X,TOT}$ - diamonds), the contribution from the soft components ($F_\mathrm{X1}$ and $F_\mathrm{X2}$ - triangles), the  boundary layer plasma ($F_\mathrm{X3}$ - bullets), and that of the reflection ($F_\mathrm{ref}$ - squares). The vertical shaded area shows the approximate dates of the periastron passage (2019.9$\pm$0.1), while the horizontal dashed line represents the estimated median value of 1.4$\times10^{-11}$~erg~cm$^{-2}$~s$^{-1}$ for $F_\mathrm{X,TOT}$.
}
\label{fig:time_flux}
\end{figure}

We illustrate in Fig.~\ref{fig:time_flux} the variation of the total flux $F_\mathrm{X,TOT}$ and that of the different components ($F_\mathrm{X1}, F_\mathrm{X2}, F_\mathrm{X1}$ and $F_\mathrm{ref}$) during the 22 yr of evolution of R Aqr.

\subsection{Optical emission of the accretion disc}

\begin{figure*}
\includegraphics[width=\linewidth]{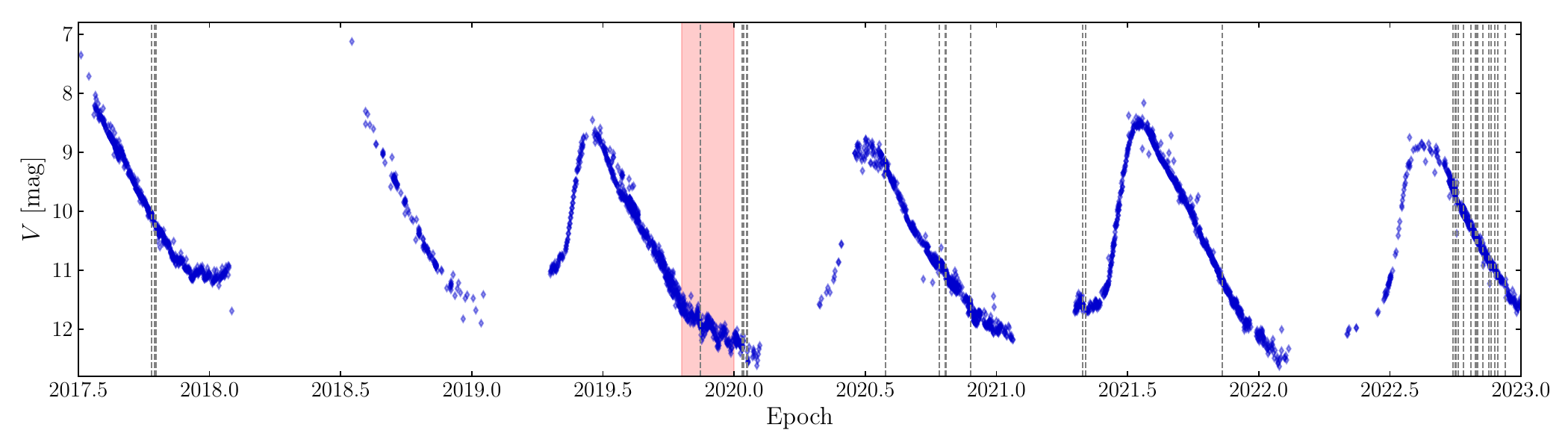}
\caption{Light curve in the Johnson $V$ band filter of R Aqr  extracted from the AAVSO database. The grey dashed vertical lines 
illustrate the dates of acquisition of the ARAS spectra and the X-ray observations. The (red) shaded region marks the periastron passage (2019.9+0.1). The figure illustrates that the observations obtained at the beginning of 2020 correspond to minimum in the pulsation of the M-type star component of R Aqr.
}
\label{fig:LC_}
\end{figure*}

The extinction-corrected, flux-calibrated ARAS spectra (see Fig.~\ref{fig:spec_opt} and Appendix~\ref{app:aras}) can be used to estimate the contribution from the accretion disc to the optical emission detected from R Aqr. The first step is to subtract the contribution from the M-type star and that of the WD component from the optical spectrum of each epoch. By fitting the absorption feature at $\lambda$=6140~\AA~ with the templates from \citet{Fluks1994}, we show that in general M8--M9 star models match the observed spectra. In contrast with the rest of the spectra, those obtained close to periastron passage (2019 Nov 15, 2020 Jan 13, 2020 Jan 14, 2020 Jan 19 and 2020 Aug 30) are best reproduced by M5--M6 star models (see the top panel in Fig.~\ref{fig:spec_opt}, and Fig.~\ref{fig:all_aras_f} and ~\ref{fig:all_aras_f2} in Appendix~\ref{app:aras}). However, there is still some emission above $\lambda> 6100$~\AA\, that can not be accounted with the model. We attribute this formation and destruction of dust in the circumstellar medium close to the M-type star. Fig.~\ref{fig:LC_} presents the $V$-magnitude light curve of R Aqr extracted from the AAVSO database\footnote{\url{https://www.aavso.org/}} that demonstrates that different optical spectra were obtained at different pulsation configurations of the M-type star.

The contribution from the accretion disc can be attributed to the excess of flux in the blueward part of the optical spectra. We estimated the contribution of the accretion disc to the optical flux ($F_\mathrm{opt}$) by subtracting the M-type star models and that of the WD component to finally integrate the residuals over the intervals $\Delta \lambda_\mathrm{int}$ defined in Table \ref{tab:opt_obs}. The upper limit of the $\Delta \lambda_\mathrm{int}$ is defined by the exact wavelength were the observed spectrum diverges from the fitted M-type+WD component, while the lower value is defined by a common value in all ARAS spectra. This process includes an interpolation between the M-type star models and the observed spectra using cubic interpolation with the \texttt{interpolate} module from the SciPy library in Python. 
This was done for all ARAS spectra and the results are illustrated in Fig.~\ref{fig:multi_epo_flux} with grey diamonds. 

The resultant flux $F_\mathrm{opt}$ for each ARAS spectrum was converted to luminosity ($L_\mathrm{opt}$), assuming a distance of 200$^{+60}_{-20}$ pc to R Aqr. The disc optical luminosity varied between 0.08 and 0.56 L$_\odot$ for epochs between 2017 Oct and 2022 Dec. $F_\mathrm{opt}$ and $L_\mathrm{opt}$ are also listed in columns 9 and 10 of Table~\ref{tab:opt_obs}, respectively.

\section{Discussion}
\label{sec:discussion}

\subsection{The different components of the X-ray spectra}

The analysis presented in the previous section helped us reveal the evolution of the global properties of R Aqr in the X-ray regime during half of its $\sim$22 yr binary period. 
In addition, it allowed us to follow in detail the behaviour of the different components. Their identification is crucial to interpret correctly the evolution of the X-ray properties of R\,Aqr. 
It is widely accepted that the heavily-extinguished plasma component (in our model denoted as $kT_3$) is the one associated with the boundary layer, the volume between the inner region of the accretion disc and the surface of the accreting WD. 
Given the superb spatial resolution of {\it Chandra} ACIS-S \citep{Kellogg2007,Sacchi2024}, the second soft $kT_2$ component has been directly associated with the presence of clumps in the X-ray-emitting jet of R Aqr.

\begin{figure}
\includegraphics[width=\linewidth]{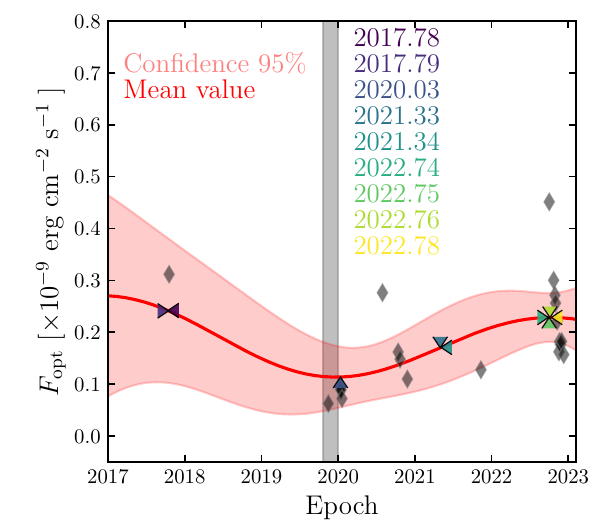}
\caption{Evolution of the optical flux of the accretion disc ($F_\mathrm{opt}$) estimated from the ARAS spectra (grey diamonds). The red line represents the mean function while the shaded red region indicates the confidence interval derived from the covariance matrix of the Gaussian process fit (see Section~\ref{sec:m_acc} for details). The coloured triangles correspond to the epochs of X-ray observations. To avoid confusion, each triangle is used as an arrow head to point to the referred value.}
\label{fig:multi_epo_flux}
\end{figure}

As for the super soft $kT_1$ component, there is some debate on its origin in $\beta/\delta$ symbiotic systems. 
Black body models have been used to fit the super soft ($E$=0.3--0.5 keV) emission from $\beta/\delta$ X-ray-emitting symbiotic systems arguing that this emission also arises as a result of the accretion process. 
Consequently, the luminosity of the super soft component is often used to estimate the mass accretion rate \citep[see for example][]{Luna2018,Sacchi2024}. 
In the case of $\beta/\delta$ sources, there are however two arguments against this hypothesis \citep[see discussion in][]{Toala2024_single}. First, the hydrogen column density component of the soft emission is more consistent with that expected from the ISM along the direction of the symbiotic system. A similar situation led \citet{Mukai2007} to suggest that the super soft component in the X-ray spectrum of CH\,Cyg should have an extended origin (outside the boundary layer region), very likely related to the jet seen from this symbiotic system \citep[][]{Galloway2004,Karovska2007}. 
Secondly, high-resolution spectra of the super soft X-ray component exhibit emission lines. For instance, in the symbiotic recurrent nova system T\,CrB, \citet{Toala2024} demonstrated that the {\it XMM-Newton} RGS spectra exhibited emission lines in the super soft range that should not be be modelled by a black body emission model. 
Consequently, we will support here the idea that the super soft component in the X-ray spectra of R Aqr comes from extended X-ray emission produced by the variable nature of jets in symbiotic systems \citep[see for example the case of V694 Mon in][]{Lucy2020}. In fact, \citet{Toala2022} demonstrated that this is the case for R\,Aqr were the super soft component is detected as an extended emission permeating the nebula associated with this symbiotic system.

There is no doubt that the presence of the 6.4 keV Fe emission line in symbiotic systems is produced by reflecting material in the vicinity of the boundary layer \citep[e.g.,][]{Eze2014}. Nevertheless, very little effort has been done in the literature in order to produce physically-driven models that fit this component and the best way to avoid this is to include a Gaussian profile to fit the 6.4 keV Fe emission line. Following the methodology by our group, here we have proposed that the reflecting material is the accretion disc. We estimated the effective Roche lobe radius of 4.7$^{+1.4}_{-0.5}$~AU which resulted to be very similar to the estimations of 5 AU presented by the analysis of near-IR data in \citet{Hinkle2022}. Consequently, our flared disc model was created with an outer radius of 5 AU. In contrast, the radius of the M-type star is estimated to be about 250 R$_\odot$ ($\lesssim$1.2 AU) and might not fill its Roche lobe to create an accretion disc. 
The formation of the accretion disc might alternatively proceed through a Bondi-Hoyle-Lyttleton process \citep{HL1939,BondiHoyle1944} or wind Roche lobe overflow mechanism \citep[][]{Podsiadlowski2007, deValBorro2009}.

Although the accretion disc should be the main source of reflection, given its larger density \citep[see, e.g.,][]{Lee2022}, other structures in its vicinity might be able to contribute to the reflection component to a lesser extent. 
For example, simulations of accreting WDs show the formation of large scale ($\gtrsim$100~AU) 3D spiral structures \citep[][]{Liu2017,deValBorro2017}. Those structures will be best included in further studies by applying the radiative-transfer code {\sc skirt} to the density structures obtained from hydrodynamical simulations.

\subsection{Consequences of the spectral analysis}

Based on the models obtained from the multiple {\it Chandra} and {\it XMM-Newton} observations of R\,Aqr, we found that during the period from 2000.69 to 2022.78, 
the total intrinsic X-ray flux $F_\mathrm{X,TOT}$ exhibited a certain degree of variability (see Fig.~\ref{fig:time_flux}), with an estimated median flux of 1.4$\times10^{-11}$ erg s$^{-1}$ cm$^{-2}$, and a median absolute deviation of 7.8$\times10^{-12}$ erg s$^{-1}$ cm$^{-2}$, without a clear increasing nor decreasing trend. 
Almost exactly the same behaviour is observed for the super soft component $F_\mathrm{X1}$, a characteristic that demonstrates that it dominates the total intrinsic flux. 
The non-detection of this component beyond 2021.33 has to be attributed to the ACIS instruments dramatic reduction of effective area in the soft energy range ($<$1.0 keV) rather than to its intrinsic decline. 
On the other hand, the second soft component $F_\mathrm{X2}$, exhibits a flux increase from the first to the last epoch of observation ranging about 2 orders of magnitude. 
Given that the $kT_2$ plasma component can be directly associated with X-ray-emitting clumps in the jet, it might be suggested that the jet activity increased with the system approaching periastron passage \citep[see][]{Sacchi2024}. 

We also report a direct correlation between the evolution of the flux of the boundary layer $F_\mathrm{X3}$ and the periastron passage of the stellar components of R\,Aqr. 
Fig.~\ref{fig:time_flux} shows that the lower $F_\mathrm{X3}$ values correspond approximately to apastron passage (about the year 2000), but it peaks at periastron passage ($\sim$2019.9). After this, the flux of the boundary layer started declining. This pattern might be related to the variation of the efficiency of the accretion process as R Aqr approaches periastron passage. We note that a similar behaviour is exhibited by the reflection component, although not exactly the same. For example, the peak of the reflection component does not corresponds exactly to the periastron passage, but a couple of years after it.

It is important to note here that the 2021 and 2022 {\it Chandra} observations were obtained with only a few days of difference among them, yet some variation is detected. 
For example, the 2021.33 and 2021.34 observations, obtained within one day, exhibit subtle but clear differences (see Table~\ref{tab:obs2} and Fig.~\ref{fig:time_flux}). 
The same can be stated about the 2022 observations. This situation shows that the X-ray emission from R Aqr is variable within time-scales of days.

\begin{figure}
\includegraphics[width=\linewidth]{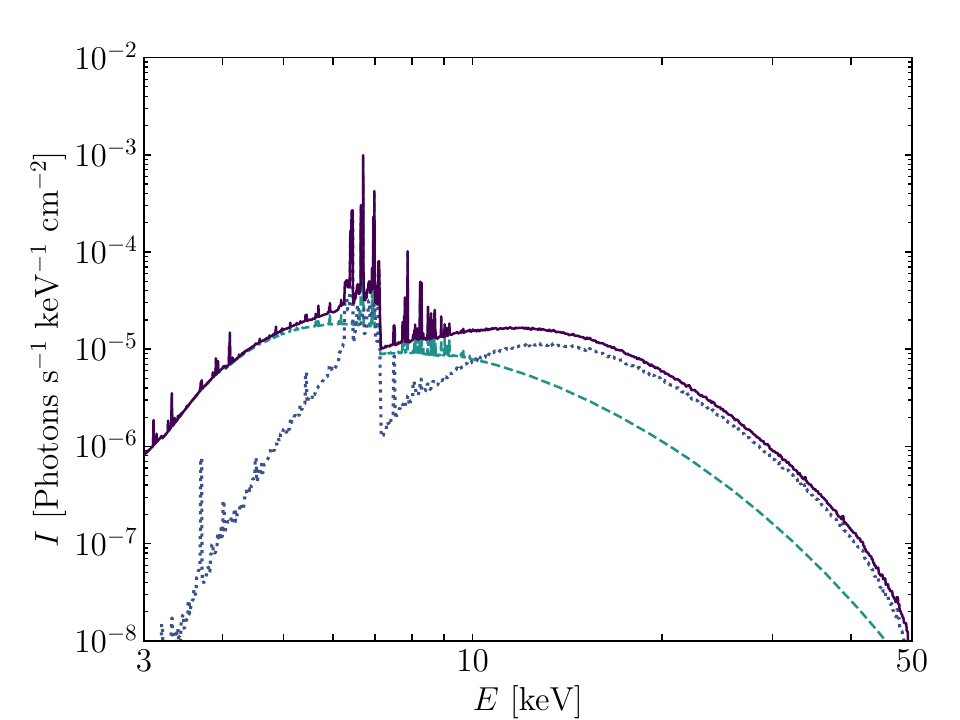}
\caption{Predictions for the hard energy range of our reflection model for the 2017.78 epoch of observations of R Aqr (solid line). The dashed and dotted lines illustrate the contribution from the plasma temperature of the boundary layer ($kT_3$) and that of reflection, respectively.}
\label{fig:nustar}
\end{figure}

Finally, we would like to remark that the {\it XMM-Newton} and {\it Chandra} observations analysed here are limited to the 0.3--10.0 keV range, but it is well-known that symbiotic systems may emit X-ray emission at higher energies, up to 100 keV \citep[e.g.,][]{Luna2019}. Previous analyses of {\it Swift} data of symbiotic systems require extremely high plasma temperatures in order to fit the high energetic part of their spectra \citep[$kT\lesssim$30~keV;][]{Kennea2009}. Nevertheless, recent analysis of multi-epoch observations of the symbiotic recurrent nova system T CrB including reflection from an accretion disc do not require such high plasma temperatures. According to the models presented in \citet{Toala2024}, reflection dominates the 15--50 keV energy range in T CrB. We note that such spectra resemble those emitted by AGN in which a Compton shoulder is one of the main components \citep[e.g.,][]{Kaspi2002}. However, in the case of symbiotic systems this feature is purely modelled by the combination of the plasma component from the boundary layer and the reflection. 
In Fig.~\ref{fig:nustar} we present the predictions for the 3.0--50.0 keV energy range from our best-fit model to the 2017.78 {\it Chandra} observations. This model predicts that reflection in R Aqr should dominate the 10.0--50.0 keV energy range but, unfortunately, there are no available X-ray observations in this energy range to compare with. Future observations covering this energy range, such as those provided by {\it NuSTAR}, are necessary in order to corroborate or improve the accuracy of the reflection models presented here.

\subsection{The evolution of $\dot{M}_\mathrm{acc}$}
\label{sec:m_acc}

Fig.~\ref{fig:multi_epo_flux} illustrates the evolution of the optical flux from the accretion disc, which is found to be somewhat constant for a time period $\sim$3 years before and after periastron passage. 

Since X-ray and optical observations are not contemporaneous, we apply a Gaussian process \citep[GP;][]{rasmussen2006} to generate a non-parametric model, which is evaluated at the X-ray epochs. This Bayesian statistical approach fits the data using a constant mean function and a covariance matrix. 
We used the Python \texttt{GPflow} library\footnote{\url{https://www.gpflow.org/}}, with a squared exponential kernel \citep[][]{wilson2013}, which we consider appropriate for describing the data correlation. By applying this model, we can sample $F_\mathrm{opt}$ at the epochs of the X-ray observations. These values are listed in Table~\ref{tab:obs2} represented by the parameter $L_\mathrm{opt}$, along with their respective uncertainties derived and propagated from the GP covariance matrix and the uncertainty in the distance. These fluxes are illustrated in Fig.~\ref{fig:multi_epo_flux} with triangles.
 
\begin{figure}
\includegraphics[width=\linewidth]{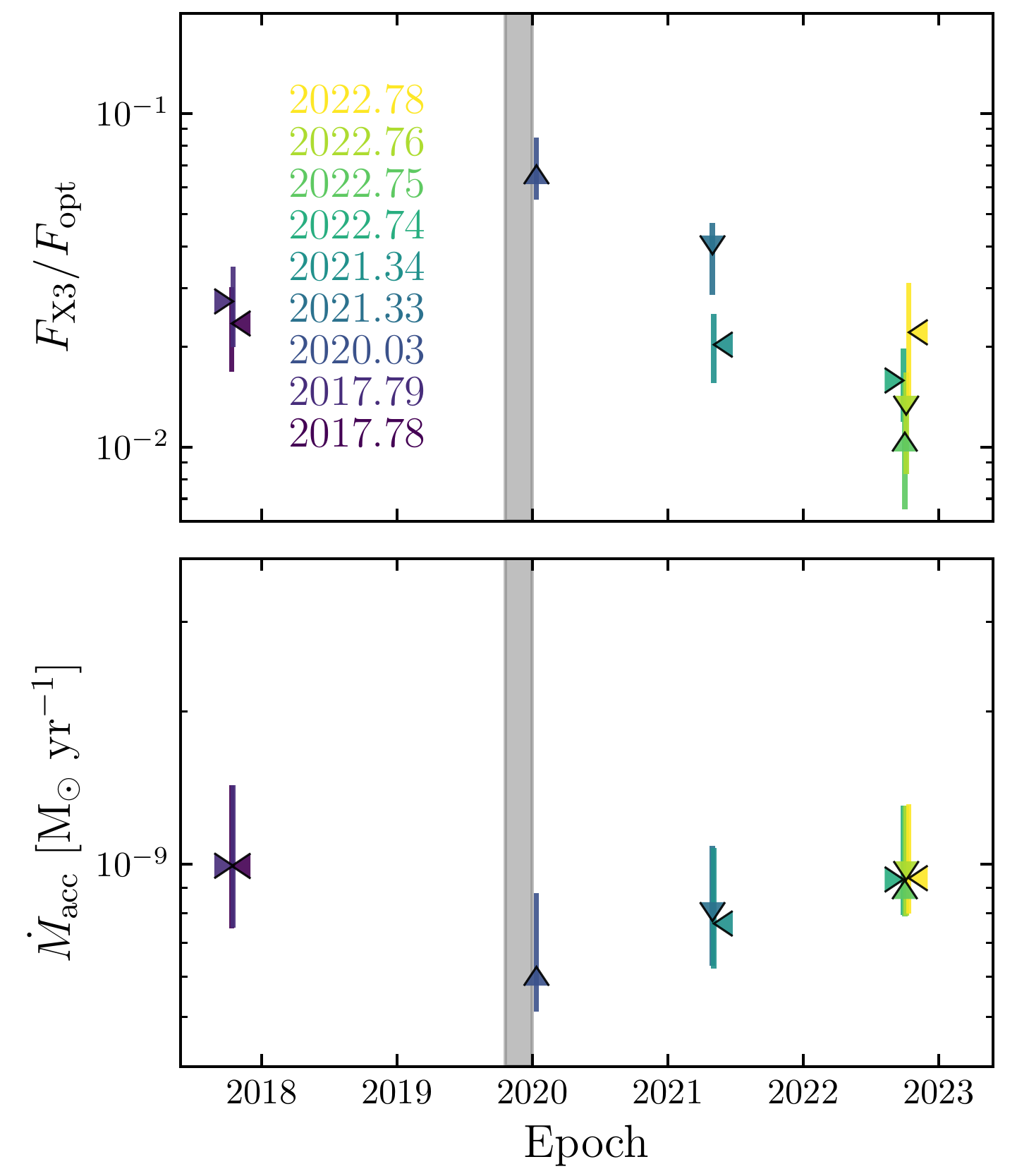}
\caption{Evolution with time of the ratio of the X-ray emission from the boundary layer $F_\mathrm{X3}$ over the optical flux of the accretion disc $F_\mathrm{opt}$ ({\it top panel}) and the mass accretion rate $\dot{M}_\mathrm{acc}$ ({\it bottom panel}). The shaded grey area represents the periastron passage (2019.9$\pm$0.1). To avoid any confusion the triangles are oriented as arrowheads, pointing to wards the value obtained for each epoch.}
\label{fig:time_flux2}
\end{figure}

Previous works have tried to correlate the optical and the X-ray emission, for example, we note the classic work of \citet{Patterson1985}, where theoretical models of accreting WDs are compared with observations. However, those works adopt the complete X-ray luminosity in order to estimate accretion mass-loss rates, but we advise against that method (see above). The ratio of the X-ray flux from the boundary layer and the optical flux from the accretion disc has values of $F_\mathrm{X3}/F_\mathrm{opt}\approx$0.03, $\approx$0.07 and $\lesssim$0.02 prior, during and after periastron passage (see top panel of Fig.~\ref{fig:time_flux2}). According to the theoretical predictions from \citet{Patterson1985}, the peak of the $F_\mathrm{X3}/F_\mathrm{opt}$ would suggest a mass accretion rate onto the WD of $\dot{M}_\mathrm{acc}\approx10^{17}$~g~s$^{-1}(\approx10^{-9}$~M$_\odot$~yr$^{-1}$).

On the other hand, assuming that the total accretion luminosity $L_\mathrm{acc}$ originates from a standard thin accretion disc \citep{SS1973,Pringle1981}, we can express it as
\begin{equation}
L_\mathrm{acc} = \frac{1}{2} \frac{G M_\mathrm{WD}}{R_\mathrm{WD}} \dot{M}_\mathrm{acc}, 
\label{eq:acc}
\end{equation}
\noindent with $G$ as the gravitational constant and $R_\mathrm{WD}$ the radius of the WD component. Where $L_\mathrm{acc}$ should be the bolometric luminosity produced by the accretion process. That is, $L_\mathrm{acc}$ should account for all the emission produced (at least) in the optical and X-ray regime. Moreover, given the fact that symbiotic systems also emit considerably in the UV \citep[see for example][and references therein]{Guerrero2024} we tried to assess the contribution of the accretion disc to this wavelength range.

Unfortunately, there are no contemporary spectroscopic UV observations of R\,Aqr. Appendix~\ref{app:disc} describes our attempt to estimate an averaged contribution from the accretion disc into the near-UV range using available {\it IUE} observations. We calculated a median spectrum for all the {\it IUE} data available in the Mikulski Archive for Space Telescopes (MAST)\footnote{\url{https://archive.stsci.edu/}}. A black body model fitted to the near-UV median spectrum resulted in $T_\mathrm{bb}$=25,000~K, $L_\mathrm{WD}=0.19^{+0.11}_{-0.04}$~L$_\odot$ and $R_\mathrm{WD}=0.025^{+0.008}_{-0.003}$~R$_{\odot}$ (see details in Appendix~\ref{app:disc}). 
After subtracting this model to the near-UV spectrum we estimate an excess of $L_\mathrm{UV}=0.07^{+0.04}_{-0.01}$~L$_\odot$ which can be attributed to the accretion disc. We note that although the different {\it IUE} near-UV spectra might suggest a certain degree of variability, the median spectrum presented in Fig.~\ref{fig:bb} seems to be a good representation of the averaged properties. 

We then proceeded to calculate the mass accretion rate using Eq.~(\ref{eq:acc}) and adopting $L_\mathrm{acc} = L_\mathrm{opt} + L_\mathrm{UV} + L_\mathrm{X3}$ with $R_\mathrm{WD}=$0.025~R$_\odot$. With these corrections, we now estimate the mass accretion rate to have values around $\dot{M}_\mathrm{acc}\approx[0.6-1]\times$10$^{-9}$~M$_\odot$~yr$^{-1}$ for the epoch interval from 2017.78 to 2022.78. The exact values are presented in Table~\ref{tab:obs2} as $\dot{M}_\mathrm{acc}$ and are illustrated in the bottom panel of Fig.~\ref{fig:time_flux2}. The lowest $\dot{M}_\mathrm{acc}$ value corresponds to the 2020.03 epoch reflecting the behaviour of the optical luminosity of the accretion disc $L_\mathrm{opt}$, which is the dominant wavelength. 

\begin{figure}
\includegraphics[width=\linewidth]{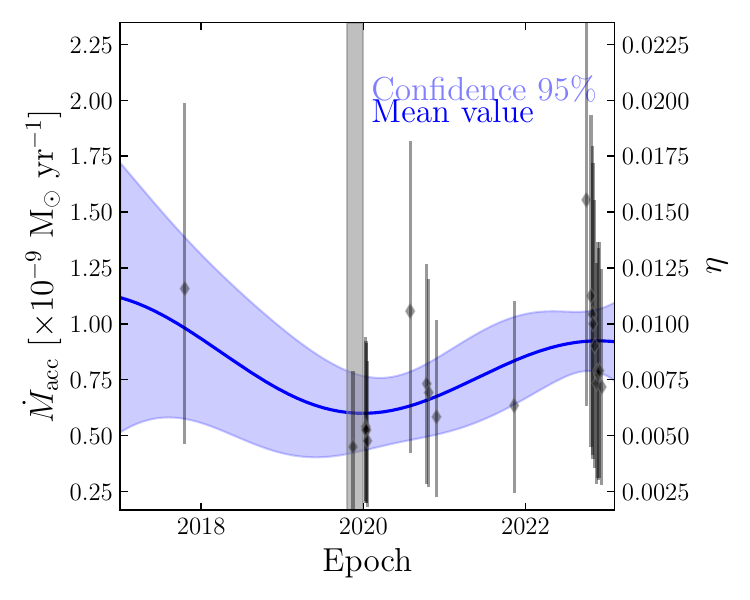}
\caption{Evolution of the mass accretion rate $\dot{M}_\mathrm{acc}$ estimated by adopting only the optical $L_\mathrm{opt}$ and UV $L_\mathrm{UV}$ emission from the accretion disc (left vertical axis) and the wind accretion efficiency $\eta$ (left vertical axis). The shaded blue region and the solid blue line result from the Gaussian process applied to the data and provide an idea of the behaviour in the intervals without data.}
\label{fig:acre_all}
\end{figure}

It is important to note here that the contribution of X-rays from the boundary layer ($L_\mathrm{X3}$) to the total bolometric luminosity produced by accretion ($L_\mathrm{acc}$) is small compared to the optical and UV fluxes of the accretion disc, given that, $L_\mathrm{X3}\approx$[10$^{-4}$--10$^{-2}$]~L$_\odot$ (see Table~\ref{tab:obs2}).
Thus, for the specific case of R Aqr we can approximate $\dot{M}_\mathrm{acc}$ by simply accounting for the optical and UV contribution from the accretion disc. That is, by assuming $L_\mathrm{acc}=L_\mathrm{opt} + L_\mathrm{UV}$. 
Consequently, in Fig.~\ref{fig:acre_all} we present the $\dot{M}_\mathrm{acc}$ estimations for all epochs of the ARAS spectra. Again, Fig.~\ref{fig:acre_all} shows the results from applying a Gaussian process to the estimated accretion rates and efficiencies derived from Eq.~(\ref{eq:acc}). The solid (blue) line represents the mean function (the most likely values given a realisation of the model), while the shaded region indicates the confidence interval associated with the covariance matrix of the Gaussian process fit. Fig.~\ref{fig:acre_all} shows that from 2017 to 2023 the mass accretion rate $\dot{M}_\mathrm{acc}$ exhibits a certain degree of variability, with a median value of 7.3$\times10^{-10}$~M$_\odot$~yr$^{-1}$.

For comparison, we mention that there are different estimations for the mass accretion rate onto the WD in R Aqr. Depending on a combination of different parameters, such as $M_\mathrm{WD}$, $R_\mathrm{WD}$, the disc temperature and the efficiency of the accreted Mira-type wind onto the WD, \citet{Burgarella1992} estimated different $\dot{M}_\mathrm{acc}$ values ranging from 1.8$\times10^{-9}$ to 3.1$\times10^{-8}$~M$_\odot$~yr$^{-1}$. 
Other works \citep[see, e.g.,][]{Henney1992,Ragland2008,Melnikov2018} assumed that $\dot{M}_\mathrm{acc}$ should be about 10 per cent or an order of magnitude lower than the mass-loss rate of the red giant component, which has been estimated to be $\dot{M}_\mathrm{Mtype}\approx10^{-7}$~M$_\odot$~yr$^{-1}$ from radio wavelengths \citep[][]{Michalitsianos1980,Spergel1983,Hollis1985}. We note that previous mass accretion rate estimates adopted a total luminosity for the boundary layer and the disc of 10~L$_\odot$, a value that is about two orders of magnitude larger than our luminosity estimates.

Fig.~\ref{fig:acre_all} also shows the evolution of the mass accretion efficiency $\eta$, which is defined as
\begin{equation}
    \eta = \frac{\dot{M}_\mathrm{acc}}{\dot{M}_\mathrm{Mtype}}.
\end{equation}
The evolution of R\,Aqr suggest efficiency values from $\eta=4.49\times10^{-3}$ up to $\eta=1.55\times10^{-2}$, with a median value of 7$\times10^{-3}$, as estimated from $\dot{M}_\mathrm{acc}$ values presented in the Table \ref{tab:opt_obs}. The minimum $\eta$ values are obtained for epochs around the periastron passage of R\,Aqr, but we attribute this to the coincidence of a pulsation minimum of the M-type star (see Fig.~\ref{fig:LC_}). 

In Appendix~\ref{app:BHL} we present an analytical model to study the accretion luminosity $L_\mathrm{acc}$ evolution for a system with the same observed properties of R Aqr. This model incorporates a reformulated Bondi-Hoyle-Lyttleton accretion mechanism, which will be comprehensively detailed and validated against other observations and simulations in \citet{Tejeda2024}. 
Our analytical model suggests that the observations can be reasonably reproduced by a M-type star with a mass-loss rate of $\dot M_\mathrm{Mtype}=10^{-7}$~M$_\odot$~yr$^{-1}$ and wind velocities between 25 and 40 km s$^{-1}$ as illustrated by the top panel of Fig.~\ref{fig:fig_BHL_luminosity} \citep[consistent with wind properties of evolved stars; see, e.g.,][]{Ramstedt2020}. 
For these wind parameters, our model predicts mass accretion efficiencies between $\eta=$ 4.77$\times10^{-3}$ and 1.64$\times10^{-2}$, consistent with the observationally derived values. The bottom panel of this figure shows similar calculations adopting different $\dot{M}_\mathrm{Mtype}$ values. This seems to suggest that the variable observed properties of R Aqr can be attributed to the variable nature of the wind properties of the Mira-type star; a property that is not incorporated in the current analytical predictions. 
The analytical solution predicts that the peak mass accretion rate occurs shortly after periastron passage. For example, with a wind velocity of 30 km/s, the maximum is reached 1.8 years after periastron, around 2021.7 (see Fig.~\ref{fig:fig_BHL_luminosity}).

\section{Summary and Conclusions}
\label{sec:conclusions}

We presented the analysis of multi-epoch {\it Chandra} ACIS-S and {\it XMM-Newton} EPIC-pn X-ray observations of the symbiotic system R\,Aqr. 
The observations cover about 22 yr of evolution of this symbiotic system, between 2000 and 2022, about half of the orbital period of the system. 
We corroborated that R\,Aqr remained a $\beta/\delta$-type X-ray-emitting symbiotic system but with a certain degree of flux variation. We analysed the X-ray spectra including a reflection model, instead of the oft-used Gaussian component to fit the 6.4 keV Fe emission line. 
This model allows us to dissect the different contributing components to the X-ray spectra: the soft extended emission from the jet, the heavily-extinguished plasma component of the boundary layer, and the contribution from reflection that naturally includes the 6.4 keV Fe fluorescent line. In addition, we analysed publicly available optical and UV data to study the properties of the accretion disc. 
All our estimates were computed adopting a distance to R\,Aqr of $d=200^{+60}_{-20}$ pc.

Our main findings can be summarised as follows:
\begin{itemize}

    \item Driven by recent description of the orbital parameters of the binary system in R Aqr, we constructed a reflecting structure with a flared disc geometry with inner and outer radii of $r_\mathrm{in}$=5$\times10^{-4}$~AU and $r_\mathrm{out}$=5 AU, respectively, and an inclination of $20^{\circ}$ with respect to the line of sight. 
    The disc is characterised by an averaged column density of $N_\mathrm{H,ref}$=5$\times10^{24}$ and a temperature of $T=10^{4}$~K. 
    The size of this disc structure is significantly smaller than the estimated radius of the M-type component, which lead us to suggest that the accretion process might be due to a Bondi-Hoyle-like process or a wind Roche lobe overflow scenario.

    \item The intrinsic total flux ($F_\mathrm{X,TOT}$) and luminosity ($L_\mathrm{X,TOT}$) computed for the 0.3--10.0 keV energy range exhibited some degree of variation, but in general we found that these varied around median values for    $F_\mathrm{X,TOT}$=1.4$\times10^{-11}$~erg~s$^{-1}$~cm$^{-2}$ and  $L_\mathrm{X,TOT}$=6.7$\times10^{31}$~erg~s$^{-1}$. On the other hand, the different components exhibit dramatic variations during the 22 yr of observations.

    \item We found that the flux from the boundary layer ($F_\mathrm{X3}$) and that of the reflection component ($F_\mathrm{ref}$) are tightly correlated to the periastron passage. They exhibit their minimum values 21 yr before periastron passage, a maximum is achieved during periastron, and then their fluxes started to decline. We predict that the fluxes of these two components will keep on declining for another $\sim$21~yr after periastron (assuming that the period of the symbiotic system is in fact 42 yr). A similar behaviour, but with larger uncertainties, is exhibited by the second soft component $kT_2$ attributed to the presence of hot X-ray-emitting pockets of gas produced by the precesing jet in R Aqr. 

    \item The super soft component ($kT_1$) dominates the total flux for epochs before 2021, then it is not detected in the last {\it Chandra} observations. We attribute this situation to the diminishing of the effective area of the {\it Chandra} ACIS-S detectors and not to an intrinsic disappearance of the super soft component. 
    This situation was corroborated by evaluating the spectra of other sources in the vicinity of R\,Aqr. 

    \item We modelled publicly available optical spectra from the ARAS database for epochs covering the 2017 Oct to 2022 Dec. The spectra were analysed to subtract the contribution from the M-type and the WD components of R Aqr to estimate the contribution from the accretion disc to the optical ($L_\mathrm{opt}$). We estimate that its optical luminosity varied between  0.08 and 0.56~L$_\odot$ during these epochs. 

    \item Using available near-UV data from the {\it IUE} satellite we estimated the properties of the WD component in R Aqr. 
    Our analysis of the median near-UV spectrum resulted in a black body component with $T_\mathrm{bb}$=25,000~K, radius $R_\mathrm{WD}$=0.025$^{+0.008}_{-0.003}$~R$_\odot$ and $L_\mathrm{WD}$=0.19$^{+0.11}_{-0.04}$~L$_\odot$. 
    This helped us estimate that the accretion disc has a contribution of $L_\mathrm{UV}$=0.07$^{+0.04}_{-0.01}$~L$_{\odot}$ in the near-UV.
    
    \item We estimated mass accretion rates $\dot{M}_\mathrm{acc}$ accounting for the optical, UV and X-ray emission produced by the accretion process, but we notice that the X-ray emission from the boundary layer does not contribute significantly to the total $L_\mathrm{acc}$. 
    Thus $\dot{M}_\mathrm{acc}$ was also computed accounting only for the optical and UV emission from the accretion disc. We found that $\dot{M}_\mathrm{acc}$ varied around a median value of 7.3$\times10^{-10}$~M$_\odot$~yr$^{-1}$. Adopting a mass-loss rate of $\dot{M}_\mathrm{Mtype}=1\times10^{-7}$~M$_\odot$~yr$^{-1}$, we were able to estimate the wind accretion efficiency, which we defined as $\eta=\dot{M}_\mathrm{acc}/\dot{M}_\mathrm{Mtype}$. This efficiency parameter resulted in values between 4.49$\times10^{-3}$ and 1.55$\times10^{-2}$ with a median value of 7$\times10^{-3}$.

    \item Our results align well with predictions from an analytical Bondi-Hoyle-Lyttleton accretion model using the same orbital parameters. 
    Notably, the observed variability in disc luminosity is consistent with the model's predictions for wind velocities between 25 and 40 km\,s$^{-1}$. Furthermore, the model predicts a mass accretion efficiency at periastron consistent with those predicted from observations.
  
    \item We made predictions for the higher energy range ($E>$10--50 keV) and noticed that the reflection component will dominate the hard X-ray emission. Future observations such as those provided by {\it NuStar} can help assessing the role of the reflection disc in R\,Aqr. 
    
\end{itemize}

\section*{Acknowledgements}
The authors thank the referee, Radoslav Zamanov, for comments and suggestions that improved our analysis and the presentation of our results. J.A.T. and D.A.V.T. acknowledge support from the UNAM PAPIIT project IN102324. D.A.V.T. thanks Consejo Nacional de Humanidades, Cientica y Tecnolog\'{i}a (CONAHCyT, Mexico) for a student grant. M.A.G.\ acknowledges financial support from grants CEX2021-001131-S funded by MCIN/AEI/10.13039/501100011033 and PID2022-142925NB-I00 from the Spanish Ministerio de Ciencia, Innovaci\'on y Universidades (MCIU) cofunded with FEDER funds. The authors thank Cesar Eduardo Feliciano Torres for the 3D model used to create the sketch of the flared disc. M.K. acknowledges the support provided via the NASA Chandra grants DD9-
20111X, GO1-22027X, GO-24014. This paper employs a list of {\it Chandra} data sets, obtained by the NASA Chandra X-ray Observatory. We acknowledge with thanks the variable star observations from the AAVSO International Database contributed by observers worldwide and used in this research. We are grateful to the ARAS team for the spectra used in this research. This work has made extensive use of NASA's Astrophysics Data System (ADS).

\section*{Data availability}
The data underlying this work are available in public archives as described in Section~\ref{sec:obs}.
The processed observations files will be shared on reasonable request to the first author.

\appendix

\section{ARAS spectra}
\label{app:aras}

\begin{figure*}
\includegraphics[width=\linewidth]{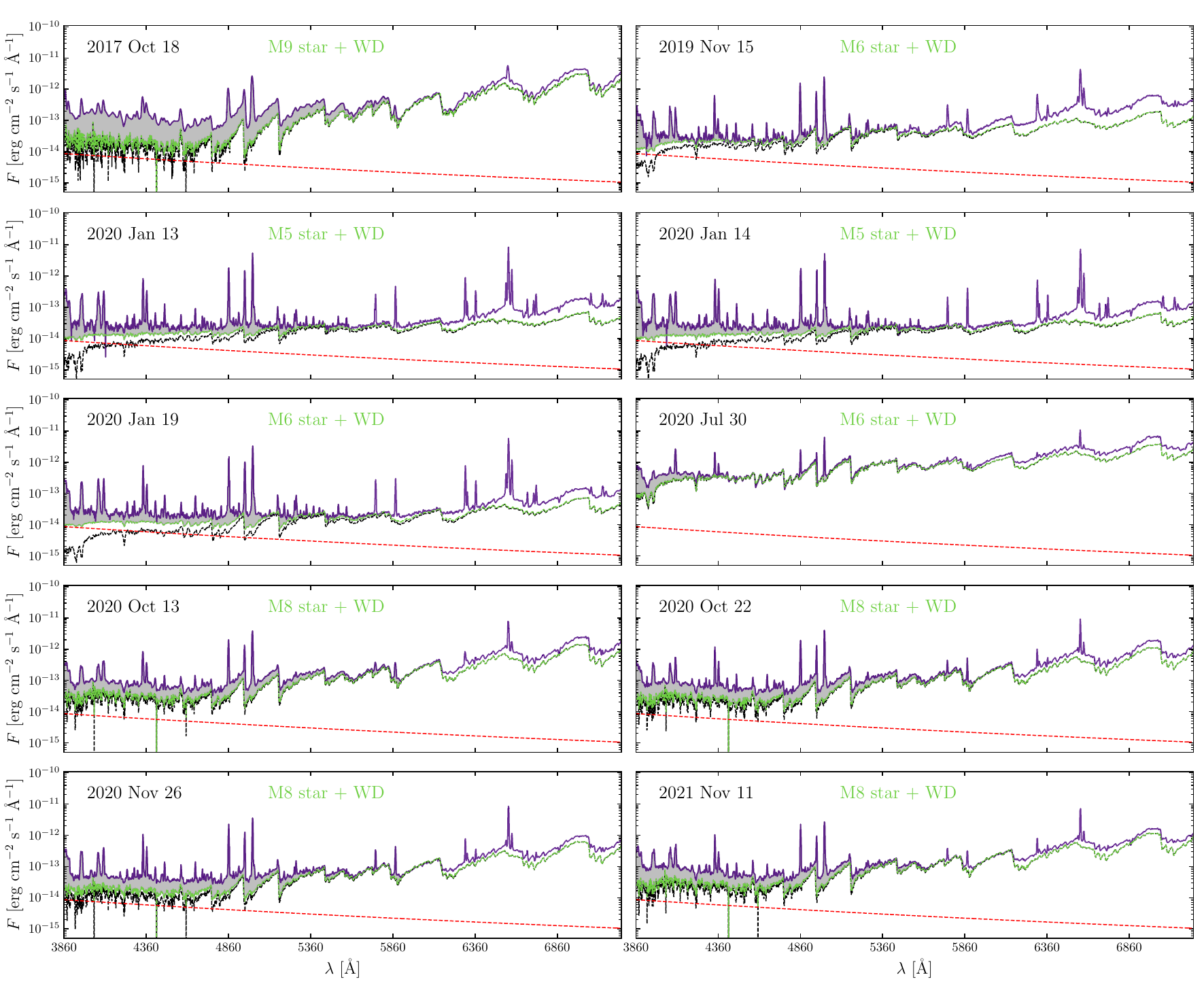}
 \caption{Extinction-corrected ARAS spectra of R~Aqr from 2017 Oct 18 to 2021 Nov 11  (solid purple line). The spectra are compared with M-type star models from \citet{Fluks1994} (black dashed lines) and the contribution from the WD with the parameters estimated in Appendix~\ref{app:disc} ($T_\mathrm{eff}=40,000$ K and $L$=0.29~L$_\odot$; red dashed line). A combined spectrum (M-type star + WD) is shown with a (green) dashed line. The gray shaded area represents the contribution of the accretion disc to the optical flux.}
\label{fig:all_aras_f}
\end{figure*}

\begin{figure*}
\includegraphics[width=\linewidth]{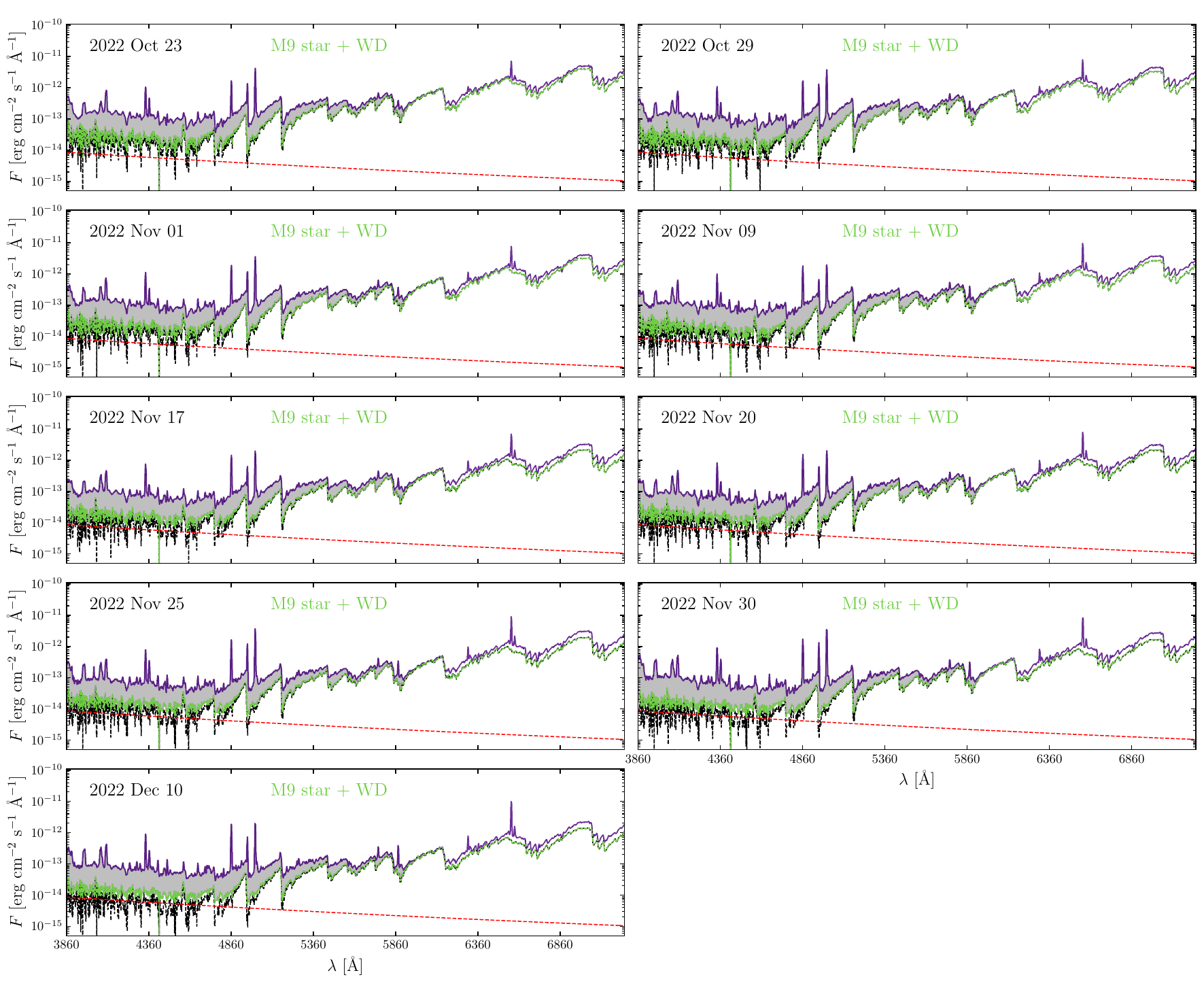}
 \caption{Same as Fig.~\ref{fig:all_aras_f} but for epochs between 2022 Oct 23 and 2022 Dec 10.}
\label{fig:all_aras_f2}
\end{figure*}

We present in Fig.~\ref{fig:all_aras_f} and \ref{fig:all_aras_f2} the extinction-corrected, flux-callibrated ARAS spectra listed in Table \ref{tab:opt_obs}. Different panels show individual spectra as well as the profiles of stellar spectra of M-type stars from \citet{Fluks1994} what were used to estimate the disc luminosity in the optical band ($L_\mathrm{opt}$). All spectra have been corrected for extinction adopting $A_\mathrm{V}=0.1$ mag and the \citet{Cardelli1998} extinction law.

\section{Ultraviolet luminosity of the accretion disc}
\label{app:disc}

In this section we present the analysis of UV data in order to estimate the contribution from the accretion disc to this wavelength. For this, we retrieved {\it IUE} spectra from R Aqr listed in the MAST with exposure times longer than 1000~s and corresponding to the low-dispersion and large aperture. These spectra cover the 1850--3350~\AA\, wavelength range.  

\begin{figure}
\includegraphics[width=\linewidth]{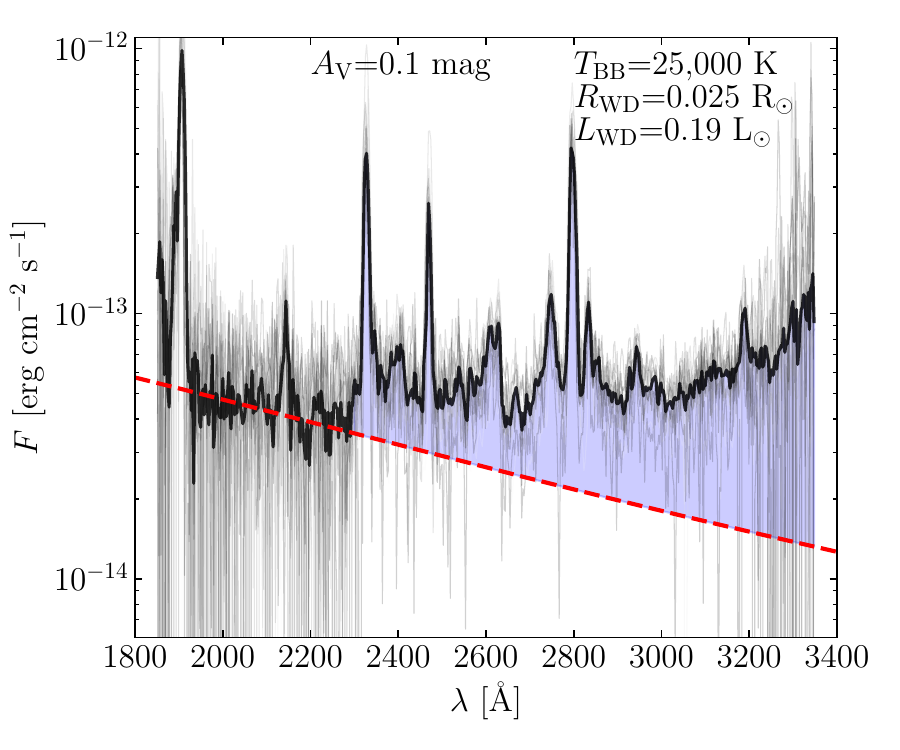}
 \caption{{\it IUE} spectrum of R Aqr (black thick line) obtained by calculating the median of all available spectra in the MAST (grey thin lines). The dashed red curve represents the contribution from a black body model with $T_\mathrm{bb}$= 25,000~K, $L_\mathrm{WD}$=0.19~L$_\odot$, and $R$=0.025~R$_\odot$. The spectrum was corrected using an extinction of $A_\mathrm{V}$=0.1~mag. The shaded region in blue represents the excess flux attributed to the contribution from the accretion disk.}
\label{fig:bb}
\end{figure}

Fig.~\ref{fig:bb} presents a median spectrum obtained from combining all available near-UV {\it IUE} spectra. To correct for reddening, we used the extinction of the interstellar medium, $A_\mathrm{V} = 0.1$ mag.

A black body spectrum was fitted to the dereddened median UV spectrum. This resulted in a model with a temperature of $T_\mathrm{BB}$=25,000 K and a bolometric luminosity of $L_\mathrm{WD}=0.19_{-0.04}^{+0.11}$ L$_\odot$ at $d=200_{-20}^{+60}$ pc. Using the Stefan-Boltzmann law we estimated the white dwarf's radius to be $R_\mathrm{WD} = 0.025^{+0.008}_{-0.003}$ R$_\odot$. This model is compared in Fig.~\ref{fig:bb} with the near-UV median spectrum of R Aqr. An excess is unambiguously detected for $\lambda>2910$~\AA.

Finally, to estimate the contributing flux from the accretion disc to the UV band, we integrated the area between the median spectrum and the black body in the interval $2296 < \lambda_\mathrm{UV} < 3350$ \AA. After converting the integrated flux, the resulting luminosity is $L_\mathrm{UV}=0.07^{+0.04}_{-0.01}$~L$_\odot$.

\section{A Bondi-Hoyle-Lyttleton accretion model for R Aqr}
\label{app:BHL}

\begin{figure}
\begin{center}
\includegraphics[width=\linewidth]{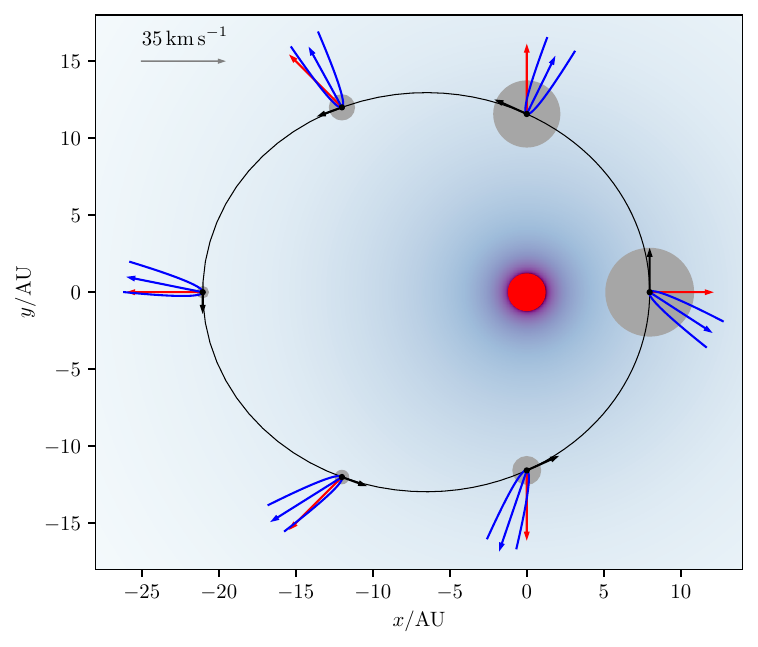}
\end{center}
\caption{This schematic illustration depicts wind accretion onto a WD in a binary system, as viewed from the primary M-type star. Black points and arrows track the WD's position and orbital velocity. Red and blue arrows represent the local wind velocity and the relative velocity between the WD and the wind, respectively. The approximate shape of the bow shock at each position is indicated by blue parabolic curves. The size of the grey circle at each point corresponds to the relative mass accretion rate at that location. The colour gradient illustrates the density distribution of the wind of the primary M-type star.}
\label{fig:BHL}
\end{figure}

In this appendix, we construct a simple analytical model to investigate wind accretion in the R Aquarii system. We assume the following orbital parameters are known:
\begin{gather}
    M_1 = 1.0\,\mathrm{M}_\odot ,\\
    M_\mathrm{WD} = 0.7\,\mathrm{M}_\odot ,\\
    e = 0.45 ,\\
    P = 42.4\,\mathrm{yr},
\end{gather}
where $M_1$  and $M_\mathrm{WD}$ represent the masses of the M-type primary and secondary WD star, $e$ is the orbital eccentricity, and $P$ is the orbital period.  Applying Kepler's third law, we derive the semi-major axis $ a = 14.51$ AU. Figure~\ref{fig:BHL} provides a schematic representation of our wind accretion model.

Following a similar approach as in \cite{Theuns1996} and \citet{Saladino18}, we employ the Bondi-Hoyle-Lyttleton (BHL) model \citep{HL1939,BondiHoyle1944} to estimate the mass accretion rate onto the white dwarf from the primary's wind as
\begin{equation}
    \Dot{M}_\text{BHL} = 4\pi \frac{(GM_\mathrm{WD})^2\rho_w}{\varv_\text{rel}^3},
    \label{dMBHL}
\end{equation}
where $\rho_w$ is the wind density and $v_\text{rel}$ represents the relative velocity between the WD and the wind. 

The primary star's stellar wind is modelled using a typical $\beta$ velocity law:
\begin{equation}
    \vec{\varv}_w(r) = \varv_\infty \left(1 - \frac{R_1}{r}\right)^\beta \,\hat{r}
\end{equation}
where $R_1 = 250$ R$_\odot \simeq 1.16$ AU is the stellar radius, $\varv_\infty$ is the terminal wind velocity and $r$ denotes the relative distance between the two stars. We adopt $\beta=2$ which best reproduces the wind profile of cool stars \citep[see][]{Lamers1999}.

Furthermore, assuming a constant mass loss rate $\dot M_\mathrm{Mtype}$, we apply the continuity equation to express the wind density as  
\begin{equation}
    \rho_w(r) = \frac{\dot M_\mathrm{Mtype}}{4\pi r^2 \varv_w} = \frac{\dot M_\mathrm{Mtype}}{4\pi a^2 \varv_\infty} \left(\frac{a}{r}\right)^2 \left(1 - \frac{R_1}{r}\right)^{-\beta}.
    \label{rhow}
\end{equation}

While, for an elliptic trajectory
\begin{equation}
    \varv_\text{rel} = \sqrt{\varv_w^2 + \varv^2_\text{WD} - 2\, \varv_w \varv_\text{WD}^r},
\end{equation}
where
\begin{gather}
    \varv_\text{WD} = \varv_0\sqrt{\frac{2a}{r} - 1},\\
    \varv_\text{WD}^r = \varv_0\frac{e}{\sqrt{1-e^2}}\sin\varphi,
\end{gather}
and 
\begin{equation}
    \varv_0 = \sqrt{G(M_1 + M_\mathrm{WD})/a}
\end{equation} is the mean orbital velocity of the WD.

As shown in more detail in \citet{Tejeda2024}, Eq.~\eqref{dMBHL} and \eqref{rhow} can be combined to calculate the mass accretion efficiency as
\begin{align}
    \eta & = \frac{\dot M_\mathrm{acc}}{\dot M_\mathrm{Mtype}} = \frac{\dot M_\mathrm{BHL}}{\dot M_\mathrm{Mtype}} \left|\frac{\varv_w - \varv_\text{WD}^r}{\varv_\text{rel}}\right|  \nonumber \\
    & = \left|1-\frac{\varv_\text{WD}^r}{\varv_w}\right|
    \left(\frac{M_\mathrm{WD}}{M_1 + M_\mathrm{WD}}\right)^2 \left(\frac{a}{r}\right)^2\left(\frac{\varv_0}{\varv_\text{rel}}\right)^4 .
\end{align}
This quantity represents the fraction of the primary's stellar wind captured by the WD star via BHL accretion. For the assumed orbital parameters of R Aqr, we find that, at pericentre ($\varphi=0$) the mass accretion efficiency ranges from $\eta = 0.005$ to 0.02 for asymptotic wind velocities between 25 and 40 km\,s$^{-1}$, respectively.

Assuming that the captured material through wind accretion ultimately forms a standard thin accretion disc around the WD, we can estimate the expected disk luminosity as \citep[][]{SS1973}
\begin{align}
    L_\mathrm{acc} & =\frac{1}{2}\frac{GM_\mathrm{WD}}{R_\mathrm{WD}} \dot{M}_\text{acc} \nonumber \\
      & \simeq 0.44 \left(\frac{\eta}{0.01}\right) \left(\frac{\dot M_\mathrm{Mtype}}{10^{-7}\ \mathrm{M_\odot\, \text{yr}^{-1}}}\right)\left(\frac{M_\mathrm{WD}}{0.7\,\mathrm{M}_\odot}\right)\left(\frac{R_\mathrm{WD}}{0.025\,\mathrm{R}_\odot}\right)^{-1} \mathrm{L}_\odot.
\end{align}

\begin{figure}
\begin{center}
\includegraphics[width=\linewidth]{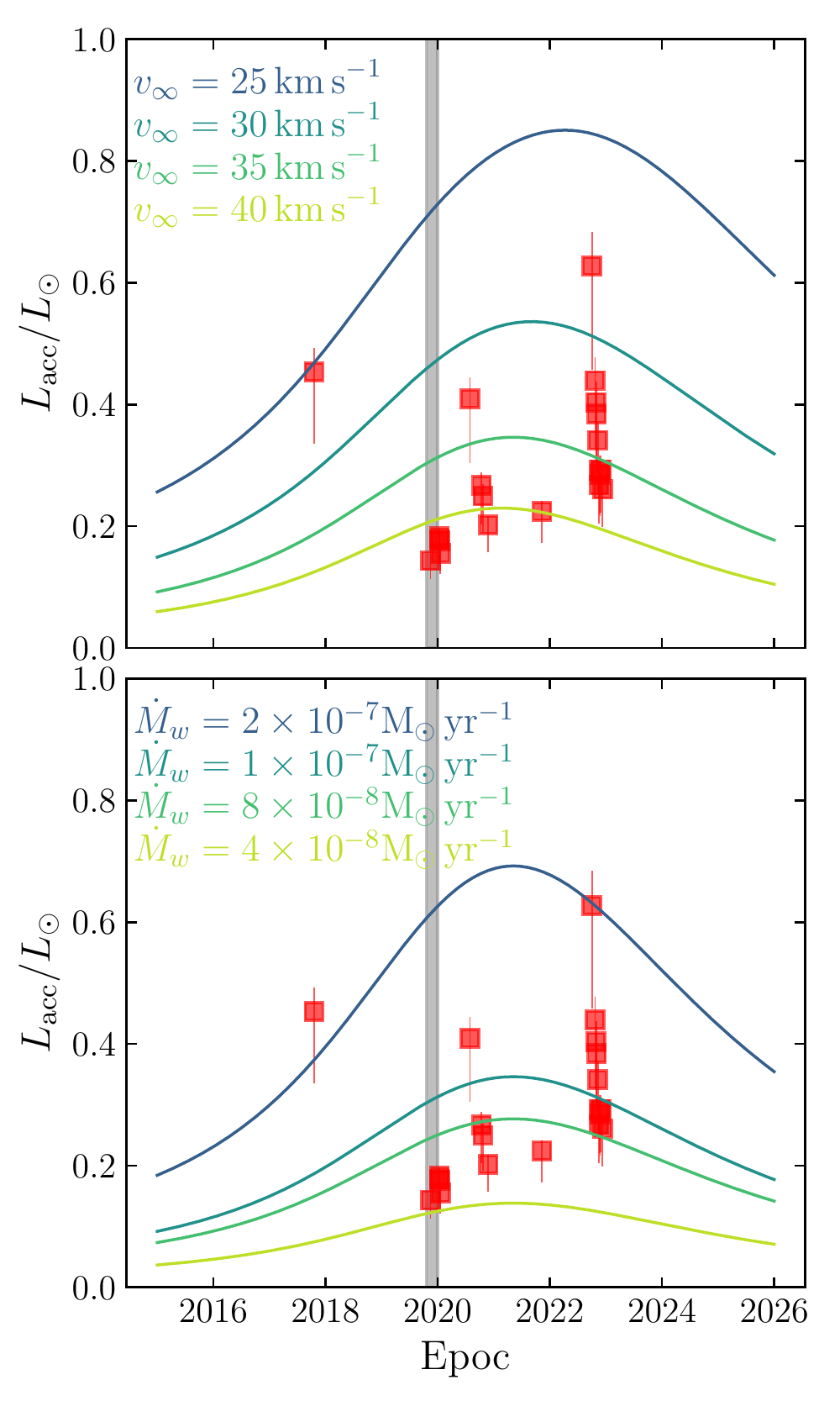}
\end{center}
\caption{Model predictions for the accreiton luminosity. (Top Panel) Four cases of $\varv_{\infty}$ are considered for a mass-loss rate of $\dot{M}_\mathrm{Mtype}$=10$^{-7}$~M$_\odot$~yr$^{-1}$ from the primary. (Bottom Panel) Results are shown for $\varv_\infty$=35~km~s$^{-1}$ and four different values of the mass loss rate $\dot{M}_\mathrm{Mtype}$. In both panels, the squares represent the values (with errors) obtained from the ARAS spectra, allowing for a direct comparison between the analytic model and observations.}
\label{fig:fig_BHL_luminosity}
\end{figure}

Figure~\ref{fig:fig_BHL_luminosity} presents the predicted accretion luminosity for four different values of the terminal wind velocity ($\varv_{\infty}$=25, 30, 35 and 45~km~s$^{-1}$) and a fixed mass loss rate from the M-type star of $\dot{M}_\mathrm{Mtype}=10^{-7}$~M$_\odot$~yr$^{-1}$. For comparison, the plot shows the results of the accretion luminosity estimated for the accretion disc in R Aqr.

The standard BHL model demonstrates broad agreement with the observational data, accurately predicting the order of magnitude of disk luminosity. This consistency is observed for the selected asymptotic wind velocities. Furthermore, for a given $\varv_\infty$, the model aligns with observations when the mass loss rate varies between 4$\times$10$^{-8}$ and 2$\times 10^{-7}$\,$\mathrm{M}_\odot$yr$^{-1}$ (see bottom panel of fig.~\ref{fig:fig_BHL_luminosity}).

The remaining discrepancies between the observations and our model can likely be attributed to the inherent simplicity of the later and the assumption of a smooth $\beta$ law for the stellar wind. In reality, the wind environment of M-type stars is expected to be more variable and/or inhomogeneous and pulsations are likely to cause abrupt variations in local wind properties.

\end{document}